\newcommand{\be}{\begin{equation}}
\newcommand{\ee}{\end{equation}}
\newcommand{\bea}{\begin{eqnarray}}
\newcommand{\eea}{\end{eqnarray}}
\newcommand{\bwt}{\begin{widetext}}
\newcommand{\ewt}{\end{widetext}}

\newcommand{\aeq}{&=&}

\newcommand{\itDelta}{{\it \Delta}}

\newcommand{\bra}{\langle}
\newcommand{\ket}{\rangle}
\newcommand{\dbra}{\bra \! \bra}
\newcommand{\dket}{\ket \! \ket}

\newcommand{\me}{\mbox{e}}

\newcommand{\rS}{{\rm S}}

\documentclass [12pt]{iopart}

\usepackage{iopams}  
\usepackage{graphicx}

\begin{document}

\title[]{Analyses of turbulence in a wind tunnel by a multifractal theory for 
probability density functions
}

\author{Toshihico Arimitsu$^1$\footnote{Corresponding 
author: arimitsu.toshi.ft@u.tsukuba.ac.jp}, Naoko Arimitsu$^2$ and Hideaki Mouri$^3$}

\address{$^1$Faculty of Pure and Applied Sciences, 
University of Tsukuba, Tsukuba, Ibaraki 305-8571, JAPAN}
\address{$^2$Faculty of Environment and Information Sciences,
Yokohama National University, Yokohama, Kanagawa 240-8501, JAPAN}
\address{$^3$Meteorological Research Institute, Tsukuba, Ibaraki 305-0052, JAPAN}

\ead{arimitsu.toshi.ft@u.tsukuba.ac.jp}

\begin{abstract}
The probability density functions (PDFs) for energy dissipation rates, created from 
time-series data of grid turbulence in a wind tunnel, are analyzed in a high precision 
by the theoretical formulae for PDFs within multifractal PDF theory
which is constructed under the assumption that there are two main elements 
constituting fully developed turbulence, i.e., coherent and incoherent elements.
The tail part of PDF, representing intermittent coherent motion, is determined 
by Tsallis-type PDF for singularity exponents essentially with one parameter
with the help of new scaling relation whose validity is checked for the case of 
the grid turbulence. 
For the central part PDF representing both contributions from 
the coherent motion and the fluctuating incoherent motion 
surrounding the former, we introduced a trial function
specified by three adjustable parameters which amazingly represent scaling behaviors 
in much wider area not restricted to the inertial range.
From the investigation of the difference between two difference formulae approximating 
velocity time-derivative, it is revealed that the connection point between the central 
and tail parts of PDF extracted by theoretical analyses of PDFs is actually the boundary of 
the two kinds of instabilities associated respectively with coherent and incoherent elements. 
\end{abstract}

\vspace{2pc}

\maketitle

\section{Introduction \label{intro}}

There are several keystone works (Mandelbrot 1974, Parisi and Frisch 1985, Benzi \etal 1984, Halsey \etal 1986, Meneveau and Sreenivasan 1987, Nelkin 1990, Hosokawa 1991, Benzi \etal 1991, She and Leveque 1994, Dubrulle 1994, She Z-S and Waymire 1995, Arimitsu T and N 2000a, 2000b, 2001, 2002, Arimitsu N and T 2002, Biferale \etal 2004, Chevillard \etal 2006)
providing the multifractal aspects for fully developed turbulence.
Only a few works (Benzi \etal 1991, Arimitsu T and N 2001, 2002, Arimitsu N and T 2002, Biferale \etal 2004, Chevillard \etal 2006)
analyze the probability density functions (PDFs) for physical quantities
representing intermittent character.
The other works 
deal with only the scaling property of the system, e.g., comparison of
the scaling exponents of velocity structure function.
Among the researches analyzing PDFs, multifractal probability density function theory 
(MPDFT) (Arimitsu T and N 2001, 2002, 2011, Arimitsu N and T 2002, 2011)
provides the most precise analysis of the fat-tail PDFs.
MPDFT is a statistical mechanical ensemble theory constructed by the authors (T.A.\ and N.A.) 
in order to analyze intermittent phenomena providing fat-tail PDFs.

To extract the intermittent character of 
the fully developed turbulence, it is necessary to have
information of self-similar hierarchical structure of the system.
This is realized by producing a series of PDFs for 
responsible singular quantities with different lengths 
\be
\ell_{n}= \ell_0 \delta^{-n}, \quad \delta>1
\quad ({n}=0,1,2,\cdots)
\label{def of delta}
\ee
that characterize the regions in which the physical quantities are coarse-grained.
The value for $\delta$ is chosen freely by observers.
Let us assume that the self-similar structure of fully developed turbulence is such 
that the choice of $\delta$ should not affect the theoretical estimation of the values for 
the fundamental quantities characterizing the turbulent system 
under consideration.
A\&A model within the framework of MPDFT itself tells us that 
this requirement is satisfied if the scaling relation has 
the form
(Arimitsu T and N 2011, Arimitsu N and T 2011)
\be
\ln 2 / (1-q)\ln \delta = 1 / \alpha_- - 1 / \alpha_+.
\label{new scaling relation}
\ee
Here, $q$ is the index associated with the R\'enyi entropy
(R\'enyi 1961) or with the Havrda-Charvat
and Tsallis
(HCT) entropy (Havrda and Charvat 1967, Tsallis 1988); 
$\alpha_\pm$ are zeros of the multifractal spectrum $f(\alpha)$
(see below in section \ref{Brief intro}).
The multifractal spectrum is uniquely related to the PDF for $\alpha$ 
(see (\ref{Tsallis prob density}) below). The PDF of $\alpha$ is related to 
the tail part of PDFs within MPDFT for those quantities revealing 
intermittent behavior whose singularity exponents can have values $\alpha < 1$, 
e.g., the energy dissipation rates, through the variable transformation 
between $\alpha$ and the physical quantities (see (\ref{epsilon alpha}) below 
for the case of the energy dissipation rates $\varepsilon_n$).
With the new scaling relation (\ref{new scaling relation}), 
observables have come to depend on the parameter $\delta$ only in the combination 
$
(1-q)\ln \delta
$.
The difference in $\delta$ is absorbed in the entropy index $q$.
\footnote{
Since almost all the PDFs that had been provided previously 
were for the case where $\delta = 2$, it has been possible 
to analyze PDFs
(Arimitsu T and N 2001, 2002, Arimitsu N and T 2002) with the scaling relation
$
1/(1-q) = 1/\alpha_- - 1/\alpha_+
\label{Tsallis scaling relation}
$
proposed by 
Costa \etal (1998) and Lyra and Tsallis (1998)
in connection with the $2^\infty$ periodic orbit. 
The orbit having the marginal instability of zero Liapunov exponent appears 
at the threshold to chaos via a period-doubling bifurcation in one-dimensional 
dissipative maps.
}

In the preceding papers, we analyzed PDFs for energy transfer rates 
(Arimitsu T and N 2011) 
and PDFs for energy dissipation rates
(Arimitsu N and T 2011),
which are given in figure~11 of Aoyama \etal (2005),
with the help of the new scaling relation, 
and checked the independence of the PDFs from $\delta$. 
It was found that the adjustable parameters for the central part PDF provide us with 
$\delta$-independent scaling behaviors as functions of $r/\eta$, and that 
the scaling properties are satisfied in much wider region not restricted to inside of 
the inertial range.
However, the number of data points used in drawing the PDFs in figure~11 of Aoyama \etal (2005)
is not enough, especially, for the precise analyses of the central part of the PDFs
performed in 
Arimitsu T and N (2011)
and 
Arimitsu N and T (2011).
Therefore, we will perform, as one of the aims of the present paper, the same analyses, 
which were done for DNS, with the help of PDFs created
from wind tunnel turbulence with a higher enough resolution in order to make sure 
if the characteristics discovered previously with rather poor resolution at the central part
are correct or not.
Since we have the raw time-series data taken from wind tunnel turbulence, 
we can create PDFs for energy dissipation rates with enough resolutions 
fit to our needs.

In this paper, we analyze the PDFs for energy dissipation rates
extracted out from the time series of the velocity field 
of a fully developed turbulence which were observed 
by one of the authors (H.M.)
in his experiment conducted in a wind tunnel (Mouri \etal 2008). 
In section~\ref{Brief intro}, we present the formulae of theoretical PDFs 
within A\&A model which are necessary 
in the following sections for the analyses of PDFs obtained from 
the experimental turbulence. 
In section~\ref{pdf edr}, we analyze
the observed PDFs for energy dissipation rates 
in a high precision with the theoretical PDF within A\&A model of MPDFT,
and verify the proposed assumption related to the magnification $\delta$.
In section~\ref{sec:physical inv of the results edr}, in order to see 
what information we can extract out from the time-series data, 
we compare two different PDFs for energy dissipation rates 
created from the time series data with different approximation for temporal derivative. 
We may learn from this how to treat the central part of PDFs to derive the information 
of incoherent fluctuating motion around the coherent turbulent motion.
Summary and discussion are provided in section~\ref{summary}.

\section{Singularity exponent and PDFs for energy dissipation rates
\label{Brief intro}}

MPDFT is constructed under the assumption, following Parisi and Frisch 
(1985), that for high Reynolds number 
the singularities distribute themselves in a multifractal way in real physical space.
The singularities stem from the invariance of the Navier-Stokes (N-S) equation 
for an incompressible fluid under the scale transformation 
$
{\vec x} \rightarrow {\vec x}' = \lambda {\vec x}
$,
accompanied by the scale changes
$
{\vec u} \rightarrow {\vec u}'= \lambda^{\alpha/3} {\vec u}
$
in velocity field, 
$
t \rightarrow t' = \lambda^{1- \alpha/3} t
$ 
in time and
$
p \rightarrow p'=\lambda^{2\alpha/3} p
$
in pressure with an arbitrary real number $\alpha$,
in the limit of large Reynolds number, i.e., the contribution from the dissipation term
in N-S equation, which is proportional to the kinematic viscosity $\nu$, is negligibly small
compared with the convection term.
In treating an actual turbulent system, however, the value $\nu$ is fixed to 
a finite value unique to the material of fluid prepared for an experiment. 
We should keep in mind that the dissipation term can become
effective depending on the region under consideration, since the term breaking the invariance 
does exist, i.e., non-zero (see the discussion in the following).

The invariance under the scale transformation leads to the scaling property
\be
\varepsilon_n / \epsilon = \left(\ell_n / \ell_0 \right)^{\alpha -1}
\label{epsilon alpha}
\ee
for the energy dissipation rate $\varepsilon_n$ averaged in the regions with diameter $\ell_n$.
Here, we put $\varepsilon_0 = \epsilon$ whose value is assumed to be constant.
The energy dissipation rate becomes singular for $\alpha < 1$, i.e.,
$
\lim_{n \rightarrow \infty} \varepsilon_n 
= \lim_{n \rightarrow \infty} \ell_n^{\alpha -1}
\rightarrow \infty
$.
The degree of singularity is specified by the singularity exponent $\alpha$ 
(Parisi and Frisch 1985).

Let us consider $\alpha$ to be a stochastic variable whose PDF $P^{(n)}(\alpha)$ is given by 
the R\'enyi
or HCT
type function (Arimitsu T and N 2000a, 2000b, 2001, 2002, 2011, Arimitsu N and T 2002, 2011):
\be
P^{(n)}(\alpha) \propto \left[ 1 - (\alpha - \alpha_0)^2/(\itDelta \alpha)^2 \right]^{n/(1-q)} 
\label{Tsallis prob density}
\ee
with
$
\itDelta \alpha = \left[2X/(1-q) \ln \delta \right]^{1/2}
$.
The domain of $\alpha$ is
$\alpha_{\rm min} \leq \alpha \leq \alpha_{\rm max}$ with
$\alpha_{\rm min}$ and $\alpha_{\rm max}$ being given by
$
\alpha_{\rm min/ max} = \alpha_0 \mp \itDelta \alpha 
$.
$q$ is the entropy index.\footnote{
The function (\ref{Tsallis prob density}) is the MaxEnt PDF derived from the R{\'e}nyi entropy
or from the HCT entropy with two constraints, one is the normalization condition
and the other is a fixed $q$-variance
(Tsallis 1988).
This choice of PDF is also quite natural since the R{\'e}nyi entropy and the HCT entropy
are directly related to the generalized dimension
(Hentschel and Procaccia 1983) describing
those systems containing multifractal structures
(Grassberger 1983).
Note that for the HCT entropy the relation is given with the help of 
the $q$-exponential
(Tsallis 2001)  which is a function satisfying 
a scaling invariance
(Suyari and Wada 2006) and reduces to the ordinary exponential for 
$q \rightarrow 1$.
}
From (\ref{Tsallis prob density}), we have for $n \gg 1$ the expression
of the multifractal spectrum 
\be
f(\alpha) = 1 + \ln \left[ 1 - 
(\alpha - \alpha_0)^2 / (\itDelta \alpha )^2 \right] / (1-q)\ln \delta.
\label{Tsallis f-alpha}
\ee
The independence of $f(\alpha)$ from $n$ is interpreted as a manifestation of the existence of 
self-similar hierarchical structure responsible for the intermittent fluid motion of turbulence.

The three parameters $\alpha_0$, $X$ and $q$ appeared in $P^{(n)}(\alpha)$ are determined 
as the functions of the intermittency exponent $\mu$ with the help of the three conditions.
One is the energy conservation law 
$
\bra \varepsilon_n \ket = \epsilon
$.
Another is the definition of the intermittency exponent $\mu$, i.e.,
$
\bra (\varepsilon_n/\epsilon )^2 \ket
= (\ell_n/\ell_0 )^{-\mu}
$.
The last condition is the scaling relation (\ref{new scaling relation})
with $\alpha_\pm$ being the solution of $f(\alpha_\pm) =0$, which
is a generalization of the one introduced by Tsallis and his coworkers
(Costa \etal 1998, Lyra and Tsallis 1998) 
to which (\ref{new scaling relation}) reduces when $\delta=2$.
Here, the average $\bra \cdots \ket$ is taken with $P^{(n)}(\alpha)$.
The parameter $q$ is determined, altogether with $\alpha_0$ and $X$, 
as a function of $\mu$ only in the combination $(1-q)\ln \delta$.
The difference in $\delta$ is absorbed into the entropy index $q$,
therefore changing the zooming rate $\delta$ 
may result in picking up the different hierarchy, containing the entropy
specified by the index $q$, out of self-similar structure of turbulence.
As the parameters are dependent on $q$ only in the combination $(1-q)\ln \delta$,
we are naturally led to the replacement of $n$ in the expression 
of $P^{(n)}(\alpha)$ in (\ref{Tsallis prob density}) with 
$
n= \tilde{n} / \ln \delta 
$.
If $\tilde{n}$ does not depend on $\delta$, $P^{(n)}(\alpha)$ becomes also 
independent of $\delta$.\footnote{
The introduction of $\tilde{n}$ is intimately related to the infinitely divisible process 
(Dubrulle 1994, She and Waymire 1995).
It is confirmed by the observation in the preset paper that $\tilde{n}$ is independent of $\delta$ 
and has values of ${\cal O}(1)$ (see table~\ref{pdf edr parameters4/1}).
Then, taking the limit $\delta \rightarrow 1+$ with a fixed value of $\tilde{n}$, one has
an infinitely divisible distribution. 
A detailed investigation of A\&A model from this view point will be given elsewhere 
in the near future.}
Note that, with the new number $\tilde{n}$, $\ell_n$ introduced 
in (\ref{def of delta}) reduces to
\be
\ell_{n}= \ell_0 \me^{-\tilde{n}}.
\label{ell n tilde}
\ee

\begin{figure}[htbp]
 \hspace*{-1.5cm}
\begin{minipage}{0.34\hsize}
  \begin{center}
   \includegraphics[width=75mm]{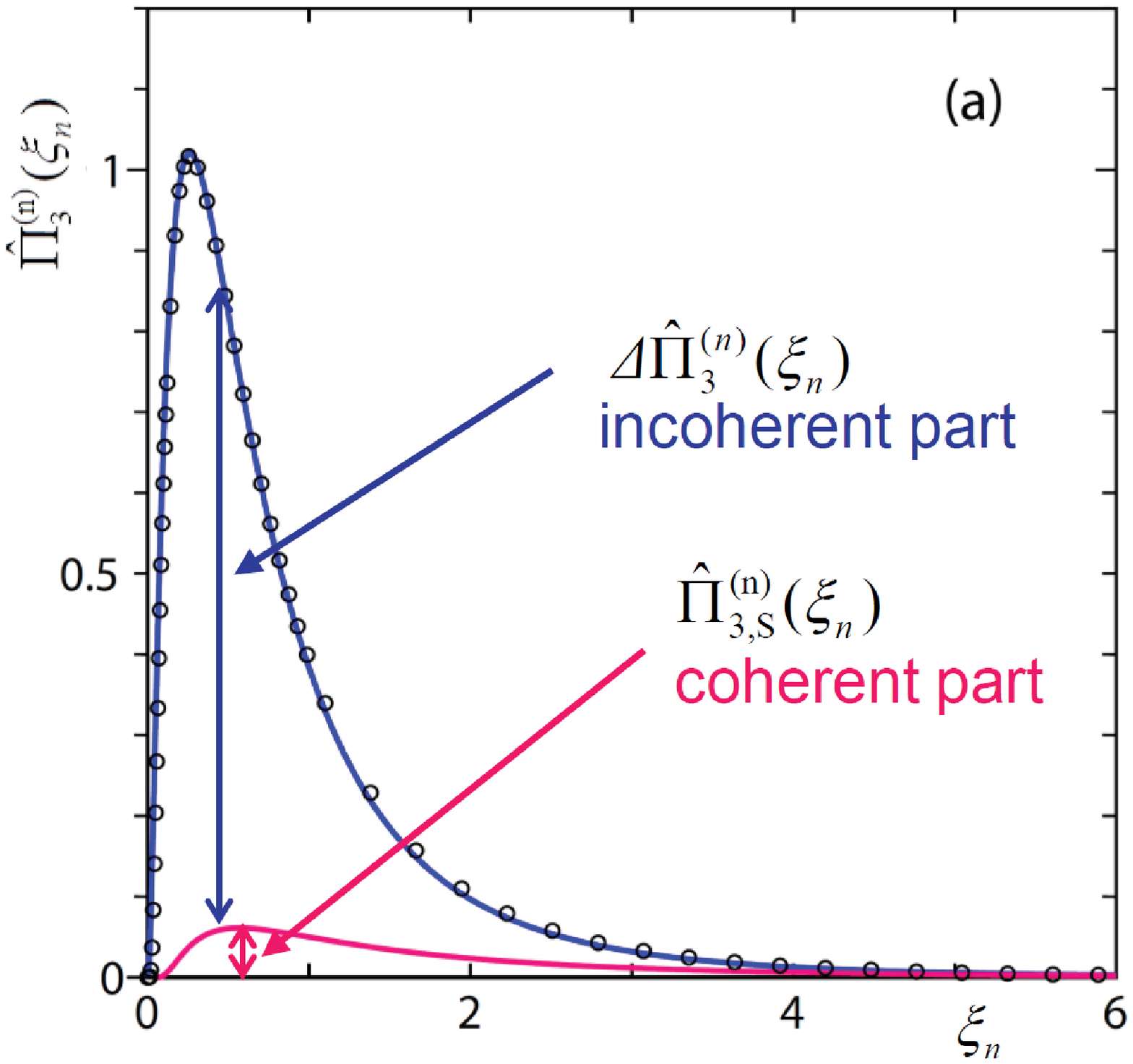}
  \end{center}
\end{minipage}
\begin{minipage}{0.33\hsize}
  \begin{center}
   \includegraphics[width=74mm]{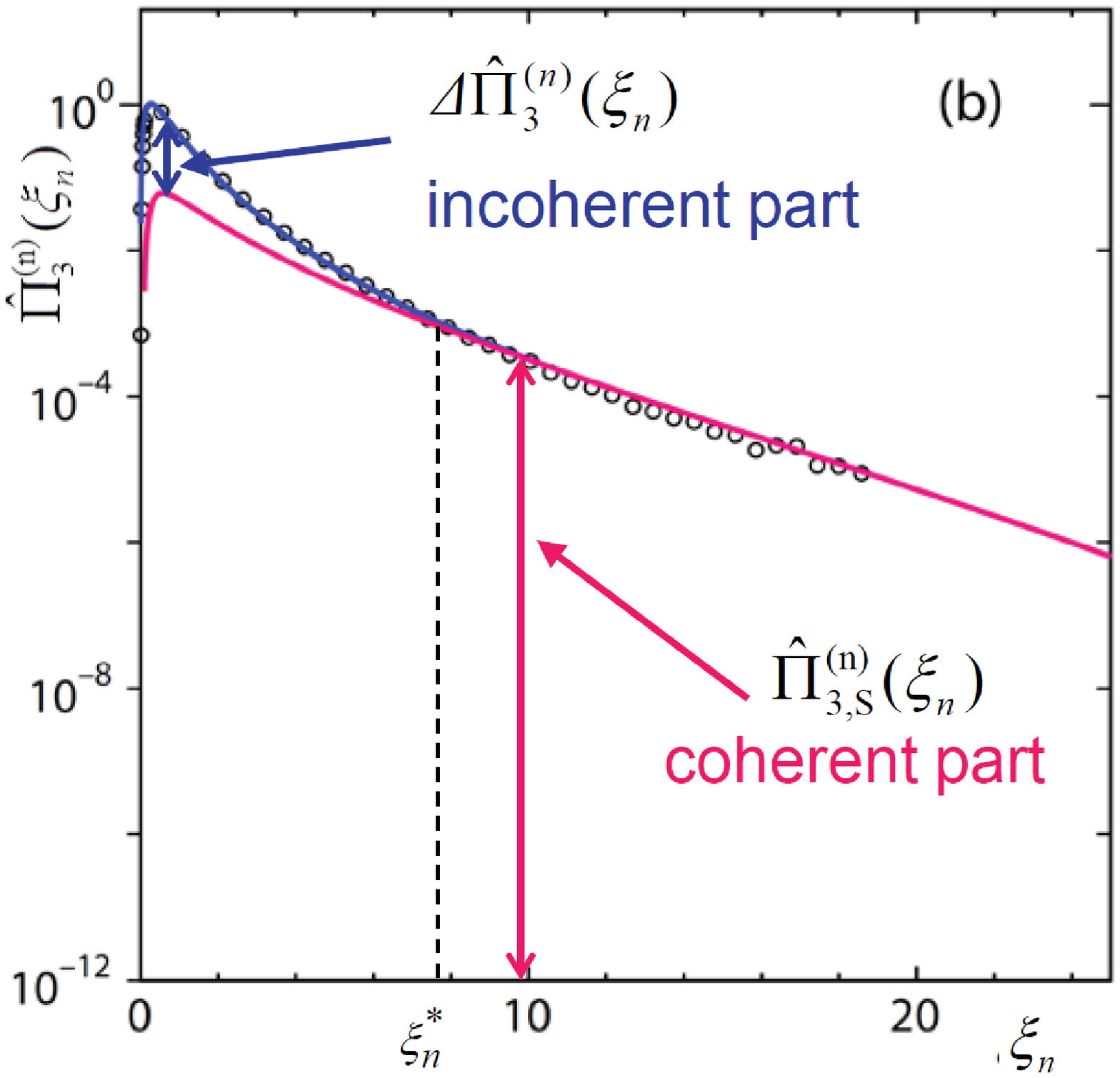}
  \end{center}
\end{minipage}
\begin{minipage}{0.33\hsize}
  \begin{center}
   \includegraphics[width=75mm]{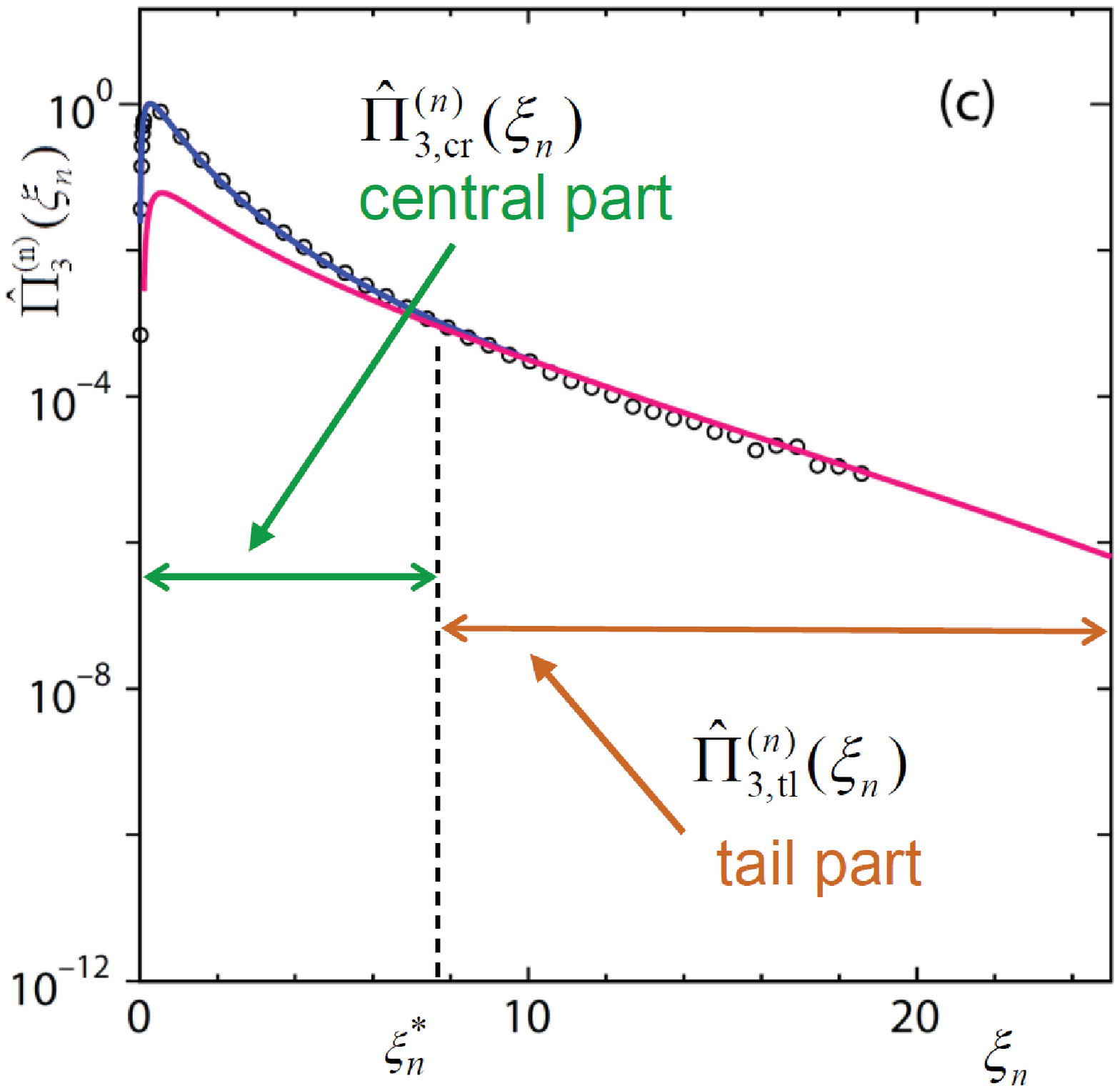}  
  \end{center}
\end{minipage}
\caption{Two kinds of divisions of PDF $\Pi_3^{(n)}(\varepsilon_n)$.
One into $\Pi^{(n)}_{3,\rS}(\varepsilon_n)$ and $\itDelta \Pi_3^{(n)}(\varepsilon_n)$ 
are given on (a) linear and (b) log scale in the vertical axes.
The other into $\hat{\Pi}_{3, {\rm cr}}^{(n)}(\xi_n)$
and $\hat{\Pi}_{3, {\rm tl}}^{(n)}(\xi_n)$ is given in (c) on log scale.
The open circles represent an experimental PDF for energy dissipation rates.
The contribution of $\itDelta \Pi_3^{(n)}(\varepsilon_n)$ to the tail part
$\hat{\Pi}_{3, {\rm tl}}^{(n)}(\xi_n)$ is negligibly small.
\label{Divisions of PDF}}
\end{figure}

MPDFT provides us with a systematic framework to make a connection between 
the PDF $P^{(n)}(\alpha)$ of the singularity exponent $\alpha$ and 
the PDF of the observed quantity such as the energy dissipation rate 
representing intermittent singular behavior in its time-evolution.
The element of fluid motion specified by the singularity exponent satisfying $\alpha <1$
takes care of the intermittent large (singular) spikes observed in the time-evolution of 
energy dissipation rate, and contributes to the tail part of PDF for energy dissipation rates
(see figure~\ref{Divisions of PDF} (a) and (b)).
This element is directly related to a coherent hierarchical structure such as 
the multi-scale Cantor set characterized by the multifractal spectrum $f(\alpha)$.
Therefore, the fluid motion controlled by this element is referred to as 
a {\em coherent} motion. 
There is another element of fluid motion due to the symmetry breaking term, i.e., 
the dissipation term in N-S equation, which produces fluctuation of fluid surrounding 
the coherent turbulent motion.
This element contributes mainly to the central part of PDF 
(see figure~\ref{Divisions of PDF} (a) and (b)).
The fluid motion provided by this element is referred to as an {\em in-coherent} motion.
Note that the central part of the PDF is constituted of two elements, 
i.e., the in-coherent and coherent motions.


Based on the above consideration, we assume that the probability 
$\Pi^{(n)}_{3}(\varepsilon_n) d\varepsilon_n$ 
can be, generally, divided into two parts as
\be
\Pi_3^{(n)}(\varepsilon_n) d\varepsilon_n = \Pi^{(n)}_{3,\rS}(\varepsilon_n) d\varepsilon_n 
+ \itDelta \Pi_3^{(n)}(\varepsilon_n) d\varepsilon_n
\label{def of Pi phi}
\ee
(see figure 1 (a) and (b)).
The first term describes the coherent motion, i.e., the contribution from the abnormal part of
the physical quantity $\varepsilon_n$ due to the fact that its singularities 
distribute themselves multifractal way in real space.
This is the part representing a coherent turbulent motion given in
the limit $\nu \rightarrow 0$ but is not equal to zero ($\nu \neq 0$).
The second term represents the contribution from the incoherent fluctuating motion.
The normalization of PDF is specified by
$
\int_0^{\infty} d\varepsilon_n  \Pi_3^{(n)}(\varepsilon_n) =1
$.
We assume that the coherent contribution is given by
 (Arimitsu T and N 2001)
$
\Pi^{(n)}_{3,\rS}(\varepsilon_n) d \varepsilon_n 
= \bar{\Pi}_{3,\rS}^{(n)} P^{(n)}(\alpha) d \alpha
$
with the variable transformation (\ref{epsilon alpha}).
For the expression of $\bar{\Pi}_{3,\rS}^{(n)}$, see Arimitsu N and T (2011).

Let us introduce another division of the PDF (see figure 1 (c)), 
i.e.,
\be
\hat{\Pi}_3^{(n)}(\xi_n) = \hat{\Pi}_{3, {\rm cr}}^{(n)}(\xi_n)
+ \hat{\Pi}_{3, {\rm tl}}^{(n)}(\xi_n),
\label{PDF cr tl}
\ee
where $\hat{\Pi}_3^{(n)}(\xi_n)$ is introduced by the relation
$\hat{\Pi}_3^{(n)}(\xi_n) d \xi_n = \Pi_3^{(n)}(\varepsilon_n) d\varepsilon_n$
with the variable transformation 
$\xi_n = \varepsilon_n / \dbra \varepsilon_n^2 \dket_{\rm c}^{1/2}$ 
where the cumulant average $\dbra \cdots \dket_{\rm c}$ is taken with 
the PDF $\Pi_3^{(n)}(\varepsilon_n)$.
The two parts of the PDF, the tail part
$\hat{\Pi}_{3, {\rm tl}}^{(n)}(\xi_n)$ 
and the central part
$\hat{\Pi}_{3, {\rm cr}}^{(n)}(\xi_n)$, 
are connected at 
$
\xi_n = \xi_n^*
$
under the conditions that they have the common value
and the common log-slope there.
Note that $\xi_n^*$ is related to $\varepsilon_n^*$ 
by $\xi_n^* = \varepsilon_n^* / \dbra \varepsilon_n^2 \dket_{\rm c}^{1/2}$
and to $\alpha^*$ by (\ref{epsilon alpha}).
The value of $\alpha^*$ is determined for each PDF as an adjusting parameter 
in the analysis of PDFs obtained by ordinary or numerical experiments.

When one creates a PDF from the time-evolution data for microscopic energy dissipation rate, 
he puts each realization into an appropriate bin according to the value $\varepsilon_n$ 
which is obtained by averaging the microscopic energy dissipation rates in each time interval 
corresponding to the length $\ell_n$.
For larger $\varepsilon_n$ values belonging to the tail part domain of the PDF, 
most of the realizations in a bin at the interval 
$\varepsilon_n \sim \varepsilon_n + d \varepsilon_n $ come from the time interval 
containing at least one intermittently large spike (singular spike) of 
microscopic energy dissipation rates. 
The bin may have negligibly small proportion of the number of realizations coming from 
those intervals with only fluctuations compared to the number of realizations
with at least one singular spike.
On the other hand, for smaller $\varepsilon_n$ values belonging to 
the central part PDF domain, the number of realizations coming from 
the time intervals containing singular spikes with smaller heights is about the same order as
the number of realizations from the time intervals containing only fluctuations, 
since the height of singular spikes contributing to this bin must be about the same height 
as fluctuations.

Under the above interpretation, it may be reasonable to assume that, for the tail part 
of PDF $\hat{\Pi}_{3, {\rm tl}}^{(n)}(\xi_n)$,
the contribution from the first term $\Pi_{3,\rS}^{(n)}(\varepsilon_n)$
in (\ref{def of Pi phi}) to the intermittent rare events dominates,
and the contribution from 
the second term $\itDelta \Pi_3^{(n)}(\varepsilon_n)$ to the events is negligible, i.e.,
\be
\hat{\Pi}_{3, {\rm tl}}^{(n)}(\xi_n)\ d \xi_n
= \Pi^{(n)}_{3,\rm S} (\varepsilon_n)\ d \varepsilon_n
\label{PDF phi large}
\ee
for $\xi_n^* \leq \xi_n$.
For $0 \leq \xi_n \leq \xi_n^*$, as there is no theory for the central part of PDF 
$\hat{\Pi}_{3, {\rm cr}}^{(n)}(\xi_n)$ at present, we put
\bea
\hat{\Pi}_{3, {\rm cr}}^{(n)}(\xi_n) d \xi_n 
\aeq \bar{\Pi}_{3}^{(n)} \me^{-[g_3(\xi_n) - g_3(\xi_n^*)]}
\ \left(\ell_n / \ell_0 \right)^{1 -f(\alpha^*)}\ 
\left(\bar{\xi}_n / \xi_n^* \right) d \xi_n
\label{PDF phi small}
\eea
with 
$
\bar{\Pi}_3^{(n)} = \bar{\Pi}_{3,S}^{(n)} 
\sqrt{\vert f^{\prime \prime}(\alpha_0) \vert / 2\pi \vert \ln (\ell_n/\ell_0) \vert} / \bar{\xi}_n
$
and a trial function of the Tsallis-type
\bea
\lefteqn{\me^{-g_3(\xi_n)} = \left(\xi_n / \xi_n^* \right)^{\theta - 1}
}
\nonumber\\
&&
\times \left\{1- \left(1-q^\prime \right)
\left[\theta+f^\prime(\alpha^*) \right]
\left[\left(\xi_n / \xi_n^* \right)^{w_3} -1 \right] / w_3 \right\}^{1/(1-q^\prime)}
\label{exp g edr}
\eea
containing minimal number of adjustable parameters, i.e., $q'$, $\theta$ and $w_3$.
The parameter $w_3$ is adjusted by the property of the experimental PDFs 
around the connection point; 
$q^\prime$ is the entropy index different from $q$ 
in (\ref{Tsallis prob density});
$\theta$ is determined by the property of PDF near $\xi_n = 0$.
For the expression of $\bar{\xi}_n$, see Arimitsu N and T (2011).
The contribution to $\hat{\Pi}_{3, {\rm cr}}^{(n)}(\xi_n)$ comes
both from coherent and incoherent motions (see figure~\ref{Divisions of PDF}).

The reason why we chose the trial function (\ref{exp g edr}) for the central part PDF 
is because it is a natural generalization of the $\chi$-square distribution function 
for the variable $y_n = (\xi_n/\xi_n^*)^{w_3}$. 
The observed value of $q'$ is in the range $1.03 \leq q' \leq 1.09$
(see table~\ref{pdf edr parameters4/1}). 
Note that in the limit $q' \rightarrow 1$ the trial function reduces to 
the $\chi$-square distribution function for $y_n$. The quantity $(\theta + w_3 -1)/w_3$ 
provides us with an estimate for the number of independent degrees of freedom 
for the dynamics contributing to the central part of PDF.

\section{Verification of assumptions through the analyses of experimental PDFs
\label{pdf edr}}

\subsection{Experimental setup and extraction of PDFs}

By means of the theoretical formula within MPDFT summarized in the last section,
we analyze PDFs of energy dissipation rates created from the time series data (Mouri \etal 2008) 
for the turbulence produced by a grid in a wind tunnel (see table~\ref{exp params}).
Measurements are performed by a hot-wire anemometer 
with a crossed-wire probe placed on the centerline of the tunnel at $4$ m downstream 
from the grid.
It is expected that turbulence around the probe is homogeneous in both 
stream-wise and span-wise directions, as the cross-section $16$ cm $\times$ $16$ cm 
of each open square surrounded by the rods constituting the grid is
small enough compared with the cross-section $3$ m $\times$ 2 m of the wind tunnel.
The cross section of a rod is $4$ cm $\times$ $4$ cm.
We also expect that the turbulence around the probe is isotropic even for larger 
scales since the values of RMS one-point velocity fluctuations for span-wise and 
stream-wise components are almost equal (see table 1).
There are still possible pitfalls about the assumption of isotropy 
(Biferale and Procaccia 2005) because of the difference of values between 
the averaged energy dissipation rates estimated with the span-wise velocity component $v$, i.e.,
$15 \nu \dbra (\partial v/\partial x)^2 \dket/2 = \dbra \varepsilon \dket =  7.98$ m$^2$ sec$^{-3}$,
and the one estimated with the stream-wise velocity fluctuation $u$, i.e.,
$15 \nu \dbra (\partial u/\partial x)^2 \dket = 8.58$ m$^2$ sec$^{-3}$
(Mouri \etal 2008).
However, as the difference is less than 10 \%, we expect that anisotropy, even if it exists,
may not affect the following analyses seriously.

\begin{table*}
\caption{Parameters of the grid turbulence in a wind tunnel (Mouri \etal 2008).
The inertial range is determined as the region where the second moment of 
the velocity differences for longitudinal component scales with the exponent $2/3$ 
with respect to the distance between the positions of two velocities used to derive
the velocity difference.
\label{exp params}}
\begin{center}
{\footnotesize
\begin{tabular}{c c} 
Quantity & Value \\
 \hline 
Microscale Reynolds number Re$_\lambda$ & 409 \\
Kolmogorov length $\eta$ & $0.138$ mm \\ 
Kinematic viscosity $\nu$ & $1.42 \times 10^{-5}$ m$^2$ sec$^{-1}$ \\
Mean velocity of downstream wind $U$ & 21.16 m sec$^{-1}$ \\
Mean energy dissipation rate 
$\dbra \varepsilon \dket = 15 \nu \dbra (\partial v/\partial x)^2 \dket/2$ & 7.98 m$^2$ sec$^{-3}$ \\
Correlation length of longitudinal velocity & $17.9$ cm \\
Inertial range & $50 \lesssim r/\eta \lesssim 150$ \\
RMS of span-wise velocity fluctuations 
$\dbra v^2 \dket^{1/2}$ & 1.06 m sec$^{-1}$ \\ 
RMS of stream-wise velocity fluctuations $\dbra u^2 \dket^{1/2}$ & 1.10 m sec$^{-1}$ \\ 
Sampling interval $\itDelta t$ & $1.43 \times 10^{-5}$ sec \\
Number of data points & $4 \times 10^8$ \\
\end{tabular}
}
\end{center}
\end{table*}

Assuming isotropy of the grid turbulence, we adopted the surrogate 
$15 \nu (\partial v/\partial x)^2/2 = 15 \nu (\partial v/\partial t)^2/2 U^2$ 
for the energy dissipation rate 
(Cleve \etal 2003, Mouri \etal 2008) with the mean velocity $U$ of downstream wind 
(see table~\ref{exp params})
where $x$-axis is chosen to the direction of the mean flow 
in a wind tunnel and $v$ is the span-wise velocity component.
Here, we used Taylor's frozen hypothesis in replacing the variable from time $t$ 
to space $x$ (see 
Mouri \etal (2008) for detail).
For the estimation of $\partial v/\partial t$, we use here the difference formula
\bea
\partial v / \partial t \simeq \itDelta^{(3)} v/\itDelta t \aeq \left\{ 8\
[v(t+\itDelta t)-v(t-\itDelta t)] 
\right. \nonumber\\
&& \left. 
- [v(t+2 \itDelta t)-v(t-2 \itDelta t)] \right\} /12 \itDelta t
\label{vel der correct up to 4th order}
\eea
where $\itDelta t$ is the sampling interval observing velocity
(see table~\ref{exp params}).
With this formula, we can have a better estimate of the velocity time derivative 
by means of $\itDelta^{(3)} v/\itDelta t$ without contamination 
up to the term of ${\cal O}(\itDelta t)^3$.
We represent the {\em local} energy dissipation rates derived 
from (\ref{vel der correct up to 4th order}) by the symbol $\varepsilon$, i.e., 
$
\varepsilon = 15 \nu (\itDelta^{(3)} v/\itDelta t)^2/2 U^2
$.

In creating the experimental PDFs for energy dissipation rates, $4 \times 10^8$ data points 
are put into $2 \times 10^4$ bins along the $\xi_n$ axis. 
We discarded those bins containing the number of data points less than 25.
Note that the average number of data points per bin is $2 \times 10^4$. 
In drawing the created PDFs for energy dissipation rates, not all the bins but
every $10^2$ bins are plotted for better visibility.
The experimental PDF in the region near the right-most end points are scattered 
because of the lower statistics due to smaller number of data points in the bins 
located there (see figure~\ref{fig:pdf edr delta_3}~(a) and 
figure~\ref{fig:pdf edr delta_3 comp 1st}~(a)).

\subsection{Analyses of experimental PDFs}

The experimental PDF is analyzed with the help of the theoretical formula for PDF
by the following procedure:
(i) Pick up three experimental PDFs with consecutive $r$ values, say, 
$r_1$, $r_2 = r_1 \delta$ and $r_3 = r_1 \delta^2$.
(ii) With a trial $\mu$ value, analyze each of the three experimental PDFs 
to find out tentative but the best values $q'$, $w_3$, $\theta$, $\alpha^*$ 
and $n_i = \ln (r_i/\ell_0) / \ln \delta$ ($i = 1$, 2, 3) for the theoretical PDF.
(iii) Check if the differences $n_3 -n_2$ and $n_2 - n_1$ are close to 1 or not.
(iv) If not, change $\mu$ value, and repeat the processes (ii) and (iii) until 
one arrives at the set of best fit parameters under the condition  
$n_3 - n_2 = n_2 - n_1 \simeq 1$ within a settled accuracy. 
(v) With thus determined common $\mu$ value, determine the best fit values 
$q'$, $w_3$, $\theta$ and $\alpha^*$ for each of other PDFs which are not 
picked out for the above processes (i) to (iv).
One notices that $n_i - n_{i-1} \simeq 1$ are satisfied automatically for every 
PDFs created from the experiment.

\begin{figure*}
\begin{center}
\resizebox*{5.5cm}{!}{\includegraphics{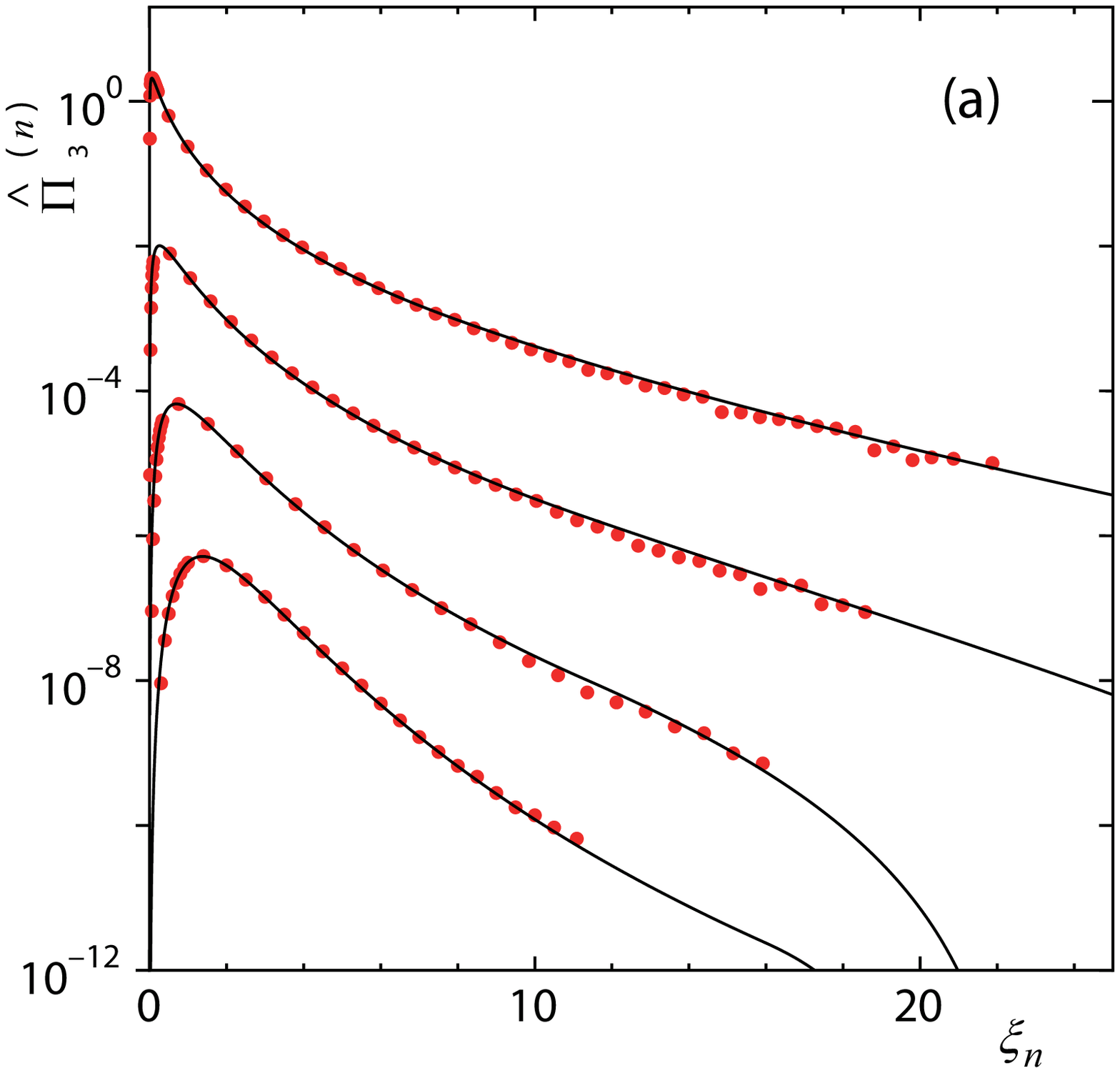}}%
\hspace*{1cm}
\resizebox*{5.65cm}{!}{\includegraphics{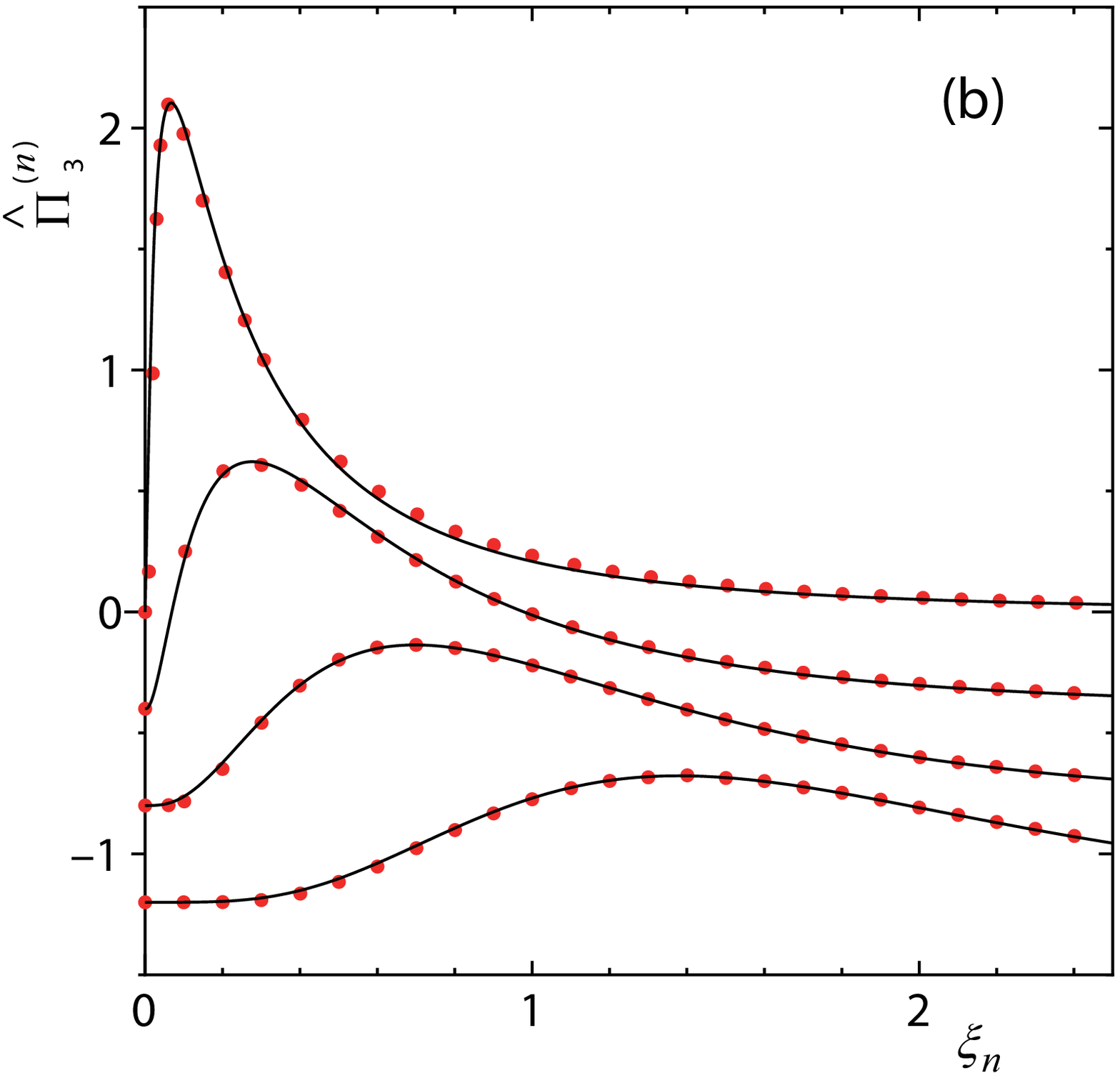}}%
\caption{PDFs of energy dissipation rates for $\delta = 3$ 
on (a) log and (b) linear scale in the vertical axes. 
For better visibility, 
each PDF is shifted by $-2$ unit along the vertical axis in (a) and by $-0.4$ 
unit along the vertical axis in (b).
Closed circles are the experimental PDFs for $r/\eta = 21.9$, $65.7$, $197$ and $591$
from the smallest value (top) to the largest value (bottom) where $r$ corresponds to $\ell_n$. 
Solid lines represent the curves given by the present theory with 
parameters listed in table~\ref{pdf edr parameters4/1}~(a).
\label{fig:pdf edr delta_3}}
\end{center}
\end{figure*}

The PDFs of energy dissipation rates are analyzed 
in figure~\ref{fig:pdf edr delta_3} for the magnification $\delta = 3$ 
on (a) log and (b) linear scale in the vertical axes.
For better visibility, 
each PDF is shifted by appropriate unit along the vertical axis.
Closed circles are the experimental data points for PDFs for the cases 
$r/\eta = 21.9$, $65.7$, $197$ and $591$ from the smallest value (top) 
to the largest value (bottom) where $r$ corresponds to $\ell_n$.
Solid lines represent the theoretical PDFs given by (\ref{PDF cr tl})
with (\ref{PDF phi large}) and (\ref{PDF phi small}).
The parameters necessary for the theoretical PDF of A\&A model are
determined as $(1-q)\ln \delta = 0.393$, $\alpha_0=1.15$ and $X=0.310$,
which turn out to be independent of $\delta$.
Other parameters are listed in table~\ref{pdf edr parameters4/1}~(a)
and table~\ref{pdf edr connection points4/1}~(a).
We performed the same analyses for other magnifications, $\delta = 2$ and 5, and
found that the extracted value $\mu = 0.260$ for the intermittency exponent
is common to three cases in which PDFs are created with the different values of magnification, 
i.e., $\delta = 2$, 3, 5.
It means that, within the analysis of the energy dissipation rates, 
the turbulent system under consideration is characterized by a unique $\mu$ value 
as it should be.

\begin{table}
\caption{Parameters of PDFs created by 
(a) the formula (\ref{vel der correct up to 4th order}) and 
(b) the formula (\ref{vel der correct up to 1st order}).
For both cases, $\mu = 0.260$ ($(1-q)\ln \delta = 0.393$, $\alpha_0=1.15$, $X=0.310$) giving $q=0.642$.
\label{pdf edr parameters4/1}}
\begin{center}
{\scriptsize
\begin{tabular}{c|c|c|c|c|c|c|c|c|c|c}
  & \multicolumn{5}{c |}{\scriptsize (a)} & \multicolumn{5}{c}{\scriptsize (b)} \\
\hline
$r/\eta$ & $n$ & $\tilde{n}$ & $q'$ & $w_3$ & $\theta$ & $n$ & $\tilde{n}$ & $q'$ & $w_3$ & $\theta$ \\ 
\hline
6.57 & 5.50 & 6.04 & 1.03 & 0.250 & 2.10 & 5.20 & 5.71 & 1.03 & 0.250 & 3.50 \\ 
21.9 &4.00 & 4.39 & 1.02 & 0.250 & 3.50 & 4.00 & 4.39 & 1.04 & 0.380 & 5.30 \\ 
65.7 & 3.00 & 3.30 & 1.05 & 0.490 & 4.10 & 3.00 & 3.30 & 1.04 & 0.450 & 5.00 \\ 
197  & 1.60 & 1.76 & 1.06 & 0.780 & 4.50 & 2.00 & 2.20 & 1.07& 0.750 & 6.00 \\ 
591  & 0.60 & 0.416 & 1.09 & 1.25 & 5.80 & 0.580 & 0.637 & 1.09 & 1.24 & 6.20 \\ 
\end{tabular}
}
\end{center}
\end{table}

\begin{table}
\caption{Connection points between the central and the tail part PDFs 
created by 
(a) the formula (\ref{vel der correct up to 4th order}) and 
(b) the formula (\ref{vel der correct up to 1st order}).
$\dbra \varepsilon \dket = 7.98$ m$^2$ sec$^{-3}$.
\label{pdf edr connection points4/1}}
\begin{center}
{\scriptsize
\begin{tabular}{c|c|c|c|c|c|c}
  & \multicolumn{3}{c |}{(a)} & \multicolumn{3}{c}{(b)} \\
\hline
$r/\eta$ & $\alpha^{*}$ & $\varepsilon_n^{*}/\dbra \varepsilon \dket$ & $\xi_n^*$ & $\alpha^{*}$ & $\varepsilon_n^{*}/\dbra \varepsilon \dket$ & $\xi_n^*$ \\ 
\hline
6.57 & 0.750 & 4.53 & 3.30 & 0.750 & 4.17 & 3.56 \\ 
21.9 & 0.700 & 3.74 & 3.56 & 0.550 & 7.22 & 5.24 \\ 
65.7 & 0.500 & 5.20 & 5.25 & 0.500 & 5.20 & 6.25 \\ 
197  & 0.280 & 3.54 & 13.6 & 0.300 & 4.66 & 13.2 \\ 
591  & 0.180 & 1.72 & 16.0 & 0.180 & 1.69 & 15.3 \\ 
\end{tabular}
}
\end{center}
\end{table}

\begin{figure}[htbp]
 \begin{minipage}{0.33\hsize}
  \begin{center}
   \includegraphics[width=51mm]{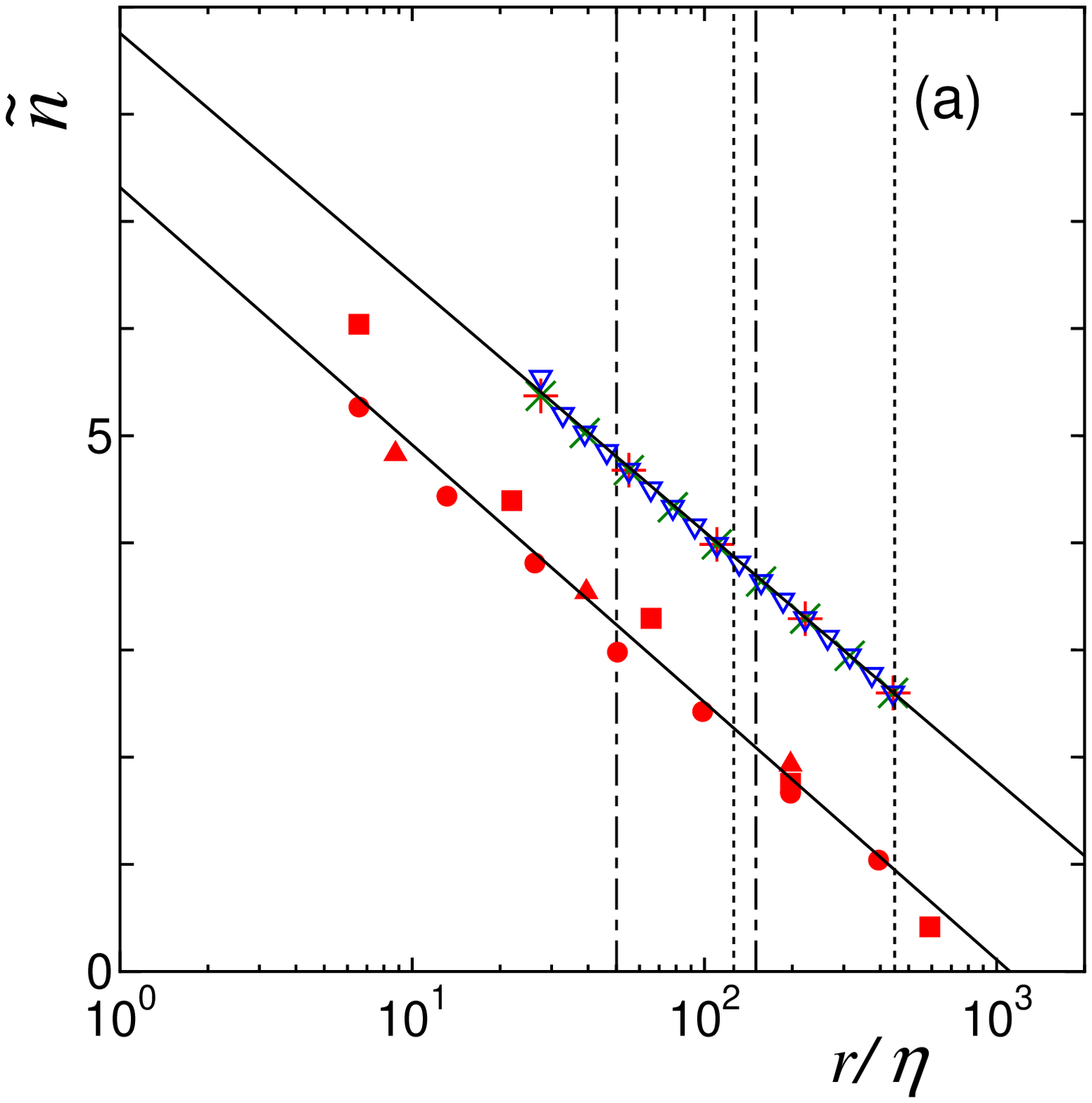}
  \end{center}
 \end{minipage}
 \begin{minipage}{0.33\hsize}
  \begin{center}
   \includegraphics[width=49mm]{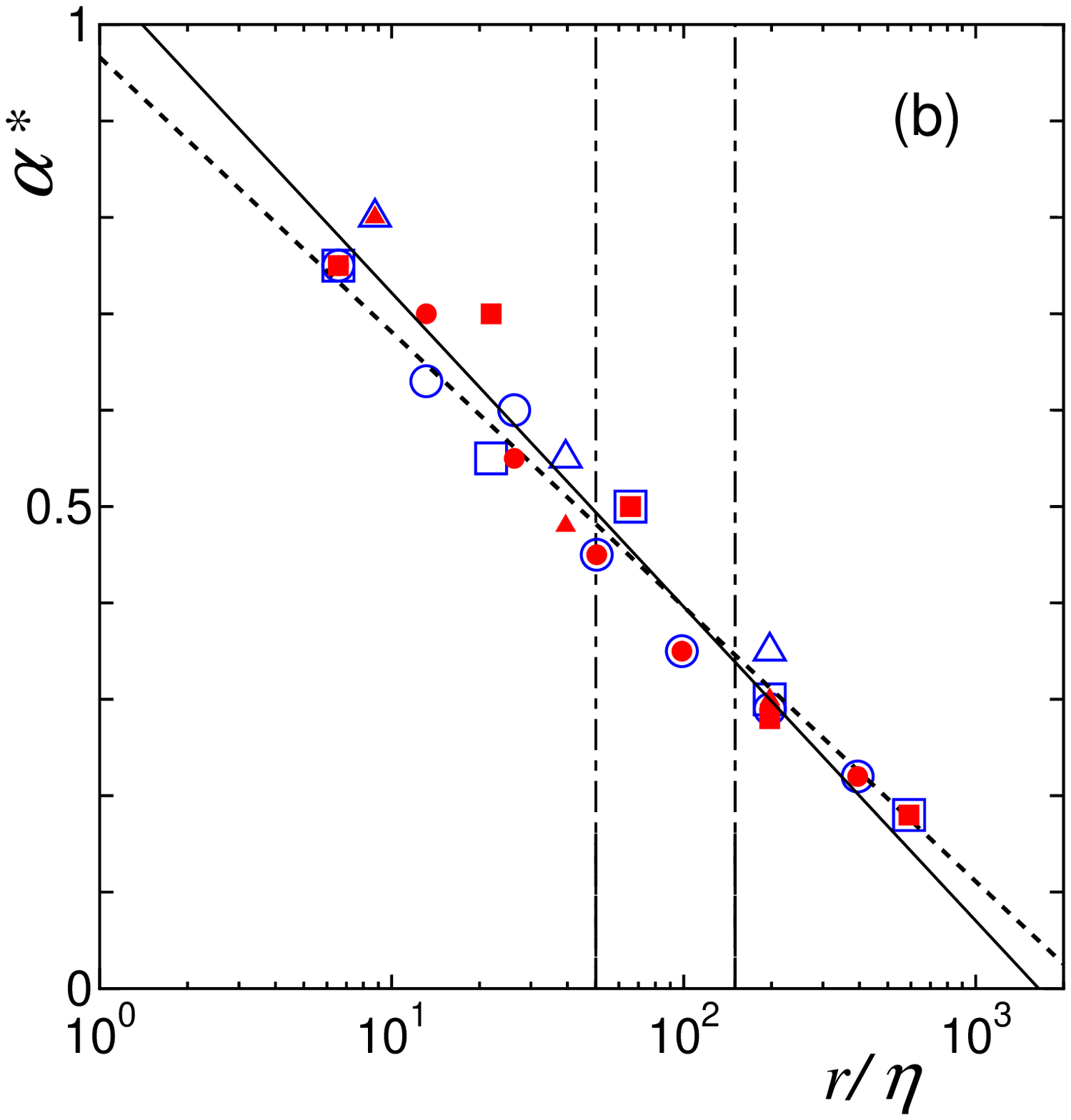}
  \end{center}
 \end{minipage}
 \begin{minipage}{0.33\hsize}
  \begin{center}
   \includegraphics[width=48mm]{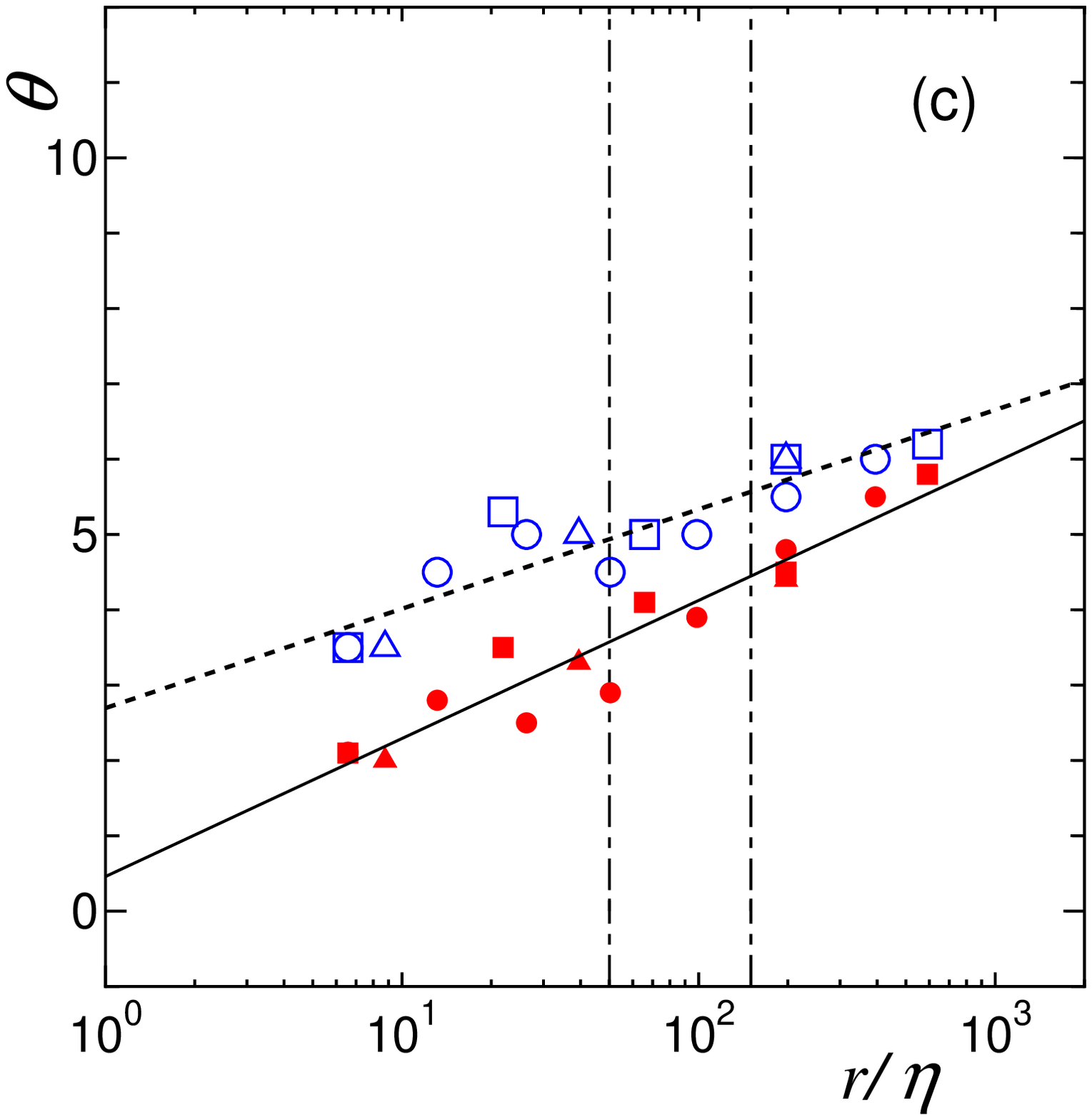}
  \end{center}
 \end{minipage}
\caption{$r/\eta$ ($=\ell_n/\eta$) dependence of (a) $\tilde{n}$, (b) $\alpha^*$ and (c) $\theta$. 
In (a), the data points extracted by the present analysis via (\ref{vel der correct up to 4th order})
are plotted by closed circles for $\delta=2$, by closed squares for $\delta=3$, 
by closed triangles for $\delta=5$, whereas those extracted by 
the DNS analysis (Arimitsu N and T 2011) are plotted by symbols nabla for $\delta=2^{1/4}$, 
by times for $\delta=2^{1/2}$, by pluses for $\delta=2$.
The empirical formula for the present (DNS) analysis is given by
 $\tilde{n}=-2.39\log_{10}(r/\eta)+7.31$
($\tilde{n}=-2.33 \log_{10}(r/\eta)+8.74$).
In (b) and (c), the data points extracted by the present analyses 
via (\ref{vel der correct up to 4th order}) (via (\ref{vel der correct up to 1st order}))
are plotted by closed (open) circles for $\delta=2$, by closed (open) squares for $\delta=3$, 
by closed (open) triangles for $\delta=5$. 
Solid (dashed) lines are the empirical formulae 
(b) $\alpha^* = - 0.326 \log_{10}(r/\eta)+1.05$ 
($\alpha^* = - 0.285 \log_{10}(r/\eta)+0.966$) and
(c) $\theta=1.83\log_{10}(r/\eta)+0.460$
($\theta=1.32\log_{10}(r/\eta)+2.70$).
The empirical formulae are obtained by using all the data points for different values of $\delta$. 
The inertial range for the present (DNS) analysis is the region between the vertical 
dash-dotted (dotted) lines.
\label{fig:pdf edr n_als_theta relation}}
\end{figure}

The dependence of $\tilde{n}$ on $r/\eta$ ($= \ell_n/\eta$) for the present analysis 
with the series of PDFs derived through (\ref{vel der correct up to 4th order})
is given in figure~\ref{fig:pdf edr n_als_theta relation} (a) 
by closed circles for $\delta = 2$, by closed squares for $\delta = 3$
and by closed triangles for $\delta = 5$.
The empirical formula for $\tilde{n}$ obtained by making use of all the data points
for $\delta = 2$, 3 and 5 with the method of least squares has the expression 
$
\tilde{n} = - 1.03 \ln (r/\eta) + 7.31
$
which is drawn by a solid line (lower line in the figure).
This proves the correctness of the assumption that 
the fundamental quantities of turbulence are independent of $\delta$.
We also include in the figure, for comparison, the data points of $\tilde{n}$ 
for $4096^3$ DNS taken from figure 4 in Arimitsu N and T (2011) and the empirical formula 
$
\tilde{n} = - 1.01 \ln (r/\eta) + 8.74
$
(upper solid line) derived with the data points by the method of least squares.
For the DNS, $\mu = 0.345$ (Arimitsu N and T 2011).
How much $\tilde{n}$ data points are scattered 
from the empirical formula (see figure~\ref{fig:pdf edr n_als_theta relation}~(a))
and also from the theoretical formula (\ref{ell n tilde}) with $\ell_n = r$ provides us with 
a measure how much we perform appropriate extraction of parameters.
The data points of $\tilde{n}$ for the turbulence in the wind tunnel are scattered more
compared with those for the turbulence in $4096^3$ DNS, as the time-series raw data 
from wind tunnel include indispensable measurement errors associated with 
readout processes, e.g., mainly, electrical noises.

The $r/\eta$ ($= \ell_n/\eta$) dependences of $\alpha^*$ and $\theta$ are given, 
respectively, in figure~\ref{fig:pdf edr n_als_theta relation} (b) and (c) 
by closed circles for $\delta = 2$, by closed squares for $\delta = 3$
and by closed triangles for $\delta = 5$,
which are extracted from the series of PDFs derived through (\ref{vel der correct up to 4th order}).
The solid line in each figure, (b) and (c), represents an empirical formula obtained from 
all the data points for $\delta = 2$, 3 and 5 by the method of least squares.
These figures prove again the correctness of the assumption that 
the fundamental quantities of turbulence are independent of $\delta$.
The value of $q^\prime$ is found to be about $q^\prime = 1.05$ 
(see table~\ref{pdf edr parameters4/1}~(a)).
We found that $w_3$ is also independent of $\delta$ and has 
a common line 
$
\log_{10} w_3 = 0.372 \log_{10} (r/\eta) + \log_{10} 0.112
$.
Note that the empirical formulae for $\tilde{n}$, $\alpha^*$ and $\theta$ 
is effective only for the region $r/\eta \gtrsim 2$ since $\theta$ should satisfy 
$\theta > 1$ (see figure~\ref{fig:pdf edr n_als_theta relation}~(c)).
We observe that the parameters $q'$, $\theta$, $w_3$ for the central part PDF and 
the connection point $\alpha^*$ have scaling behaviors in much wider region 
not restricted to inside of the inertial range.


\section{Comparison of PDFs produced with full and less contaminations
\label{sec:physical inv of the results edr}}

In this section, we analyze the PDFs for the energy dissipation rates
derived from the time-series data with the difference formula 
\be
\partial v / \partial t \simeq \itDelta^{(0)} v /\itDelta t 
= \left[ v(t+\itDelta t)-v(t) \right] / \itDelta t
\label{vel der correct up to 1st order}
\ee
in order to study what difference comes out compared with the PDFs analyzed
in section~\ref{pdf edr} which is derived by means of the difference formula 
(\ref{vel der correct up to 4th order}).
Note that the formula (\ref{vel der correct up to 1st order}) estimates the values of 
velocity time derivative with $\itDelta^{(0)} v /\itDelta t$
which may contain full contamination, i.e., from the 1st order term with respect to 
$\itDelta t$.
We introduce the symbol $\varepsilon^{(0)}$ for the {\em local} energy dissipation rates
derived from (\ref{vel der correct up to 1st order}), i.e., 
$
\varepsilon^{(0)} = 15 \nu (\itDelta^{(0)} v/\itDelta t)^2/2 U^2
$.\footnote{
We observe that $\dbra \varepsilon^{(0)} \dket = 8.08$ m$^2$ sec$^{-1}$ which is larger than 
$\dbra \varepsilon \dket$.
}
In creating the experimental PDFs for the energy dissipation rates,
we took the same procedure as used in section~\ref{pdf edr}.

\begin{figure*}
 \begin{minipage}{0.33\hsize}
  \begin{center}
   \includegraphics[width=51mm]{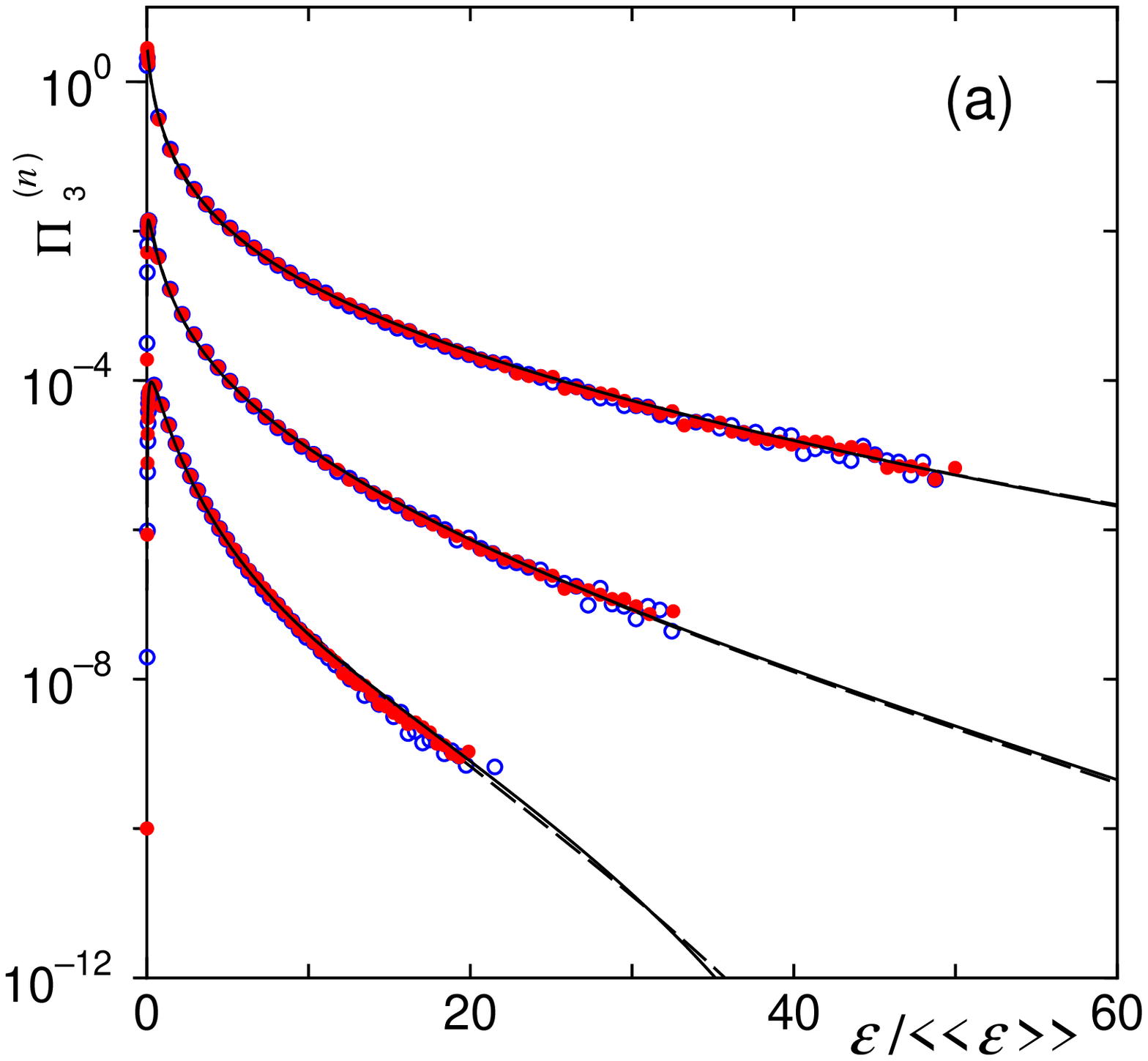}
  \end{center}
 \end{minipage}
 \begin{minipage}{0.33\hsize}
  \begin{center}
   \includegraphics[width=50mm]{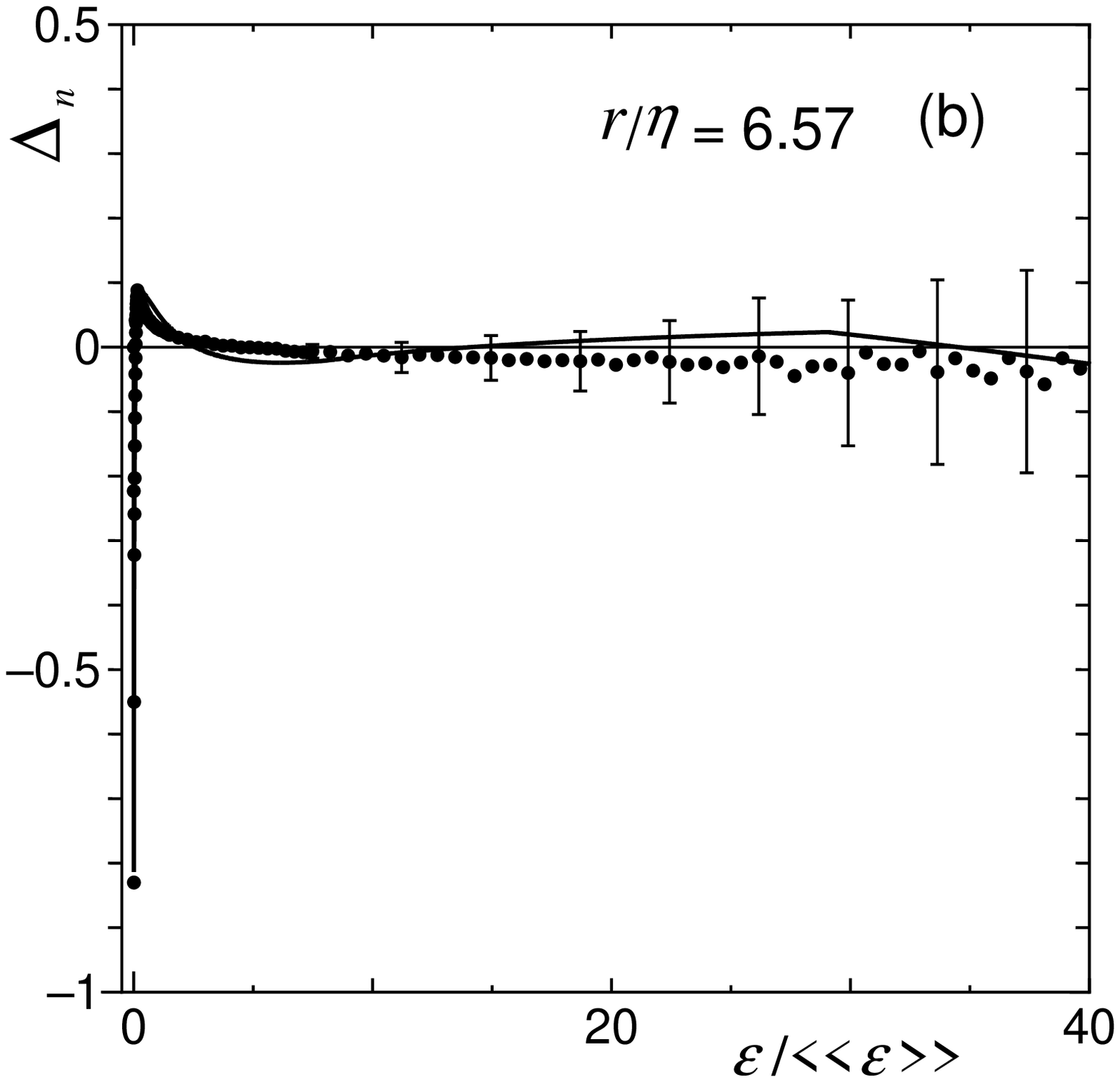}
  \end{center}
 \end{minipage}
 \begin{minipage}{0.34\hsize}
  \begin{center}
   \includegraphics[width=51mm]{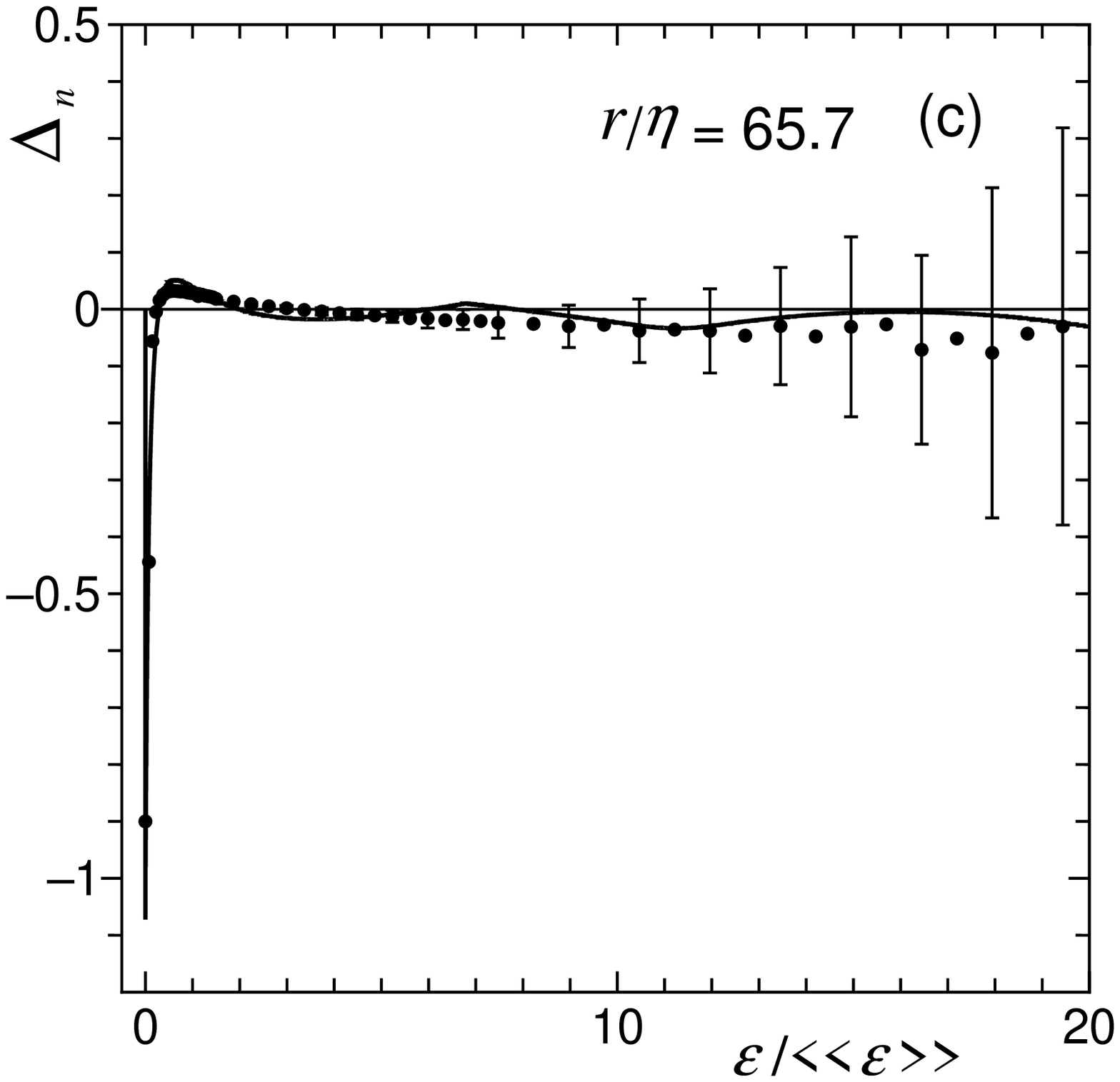}
  \end{center}
 \end{minipage}
 \begin{minipage}{0.33\hsize}
  \begin{center}
   \includegraphics[width=50mm]{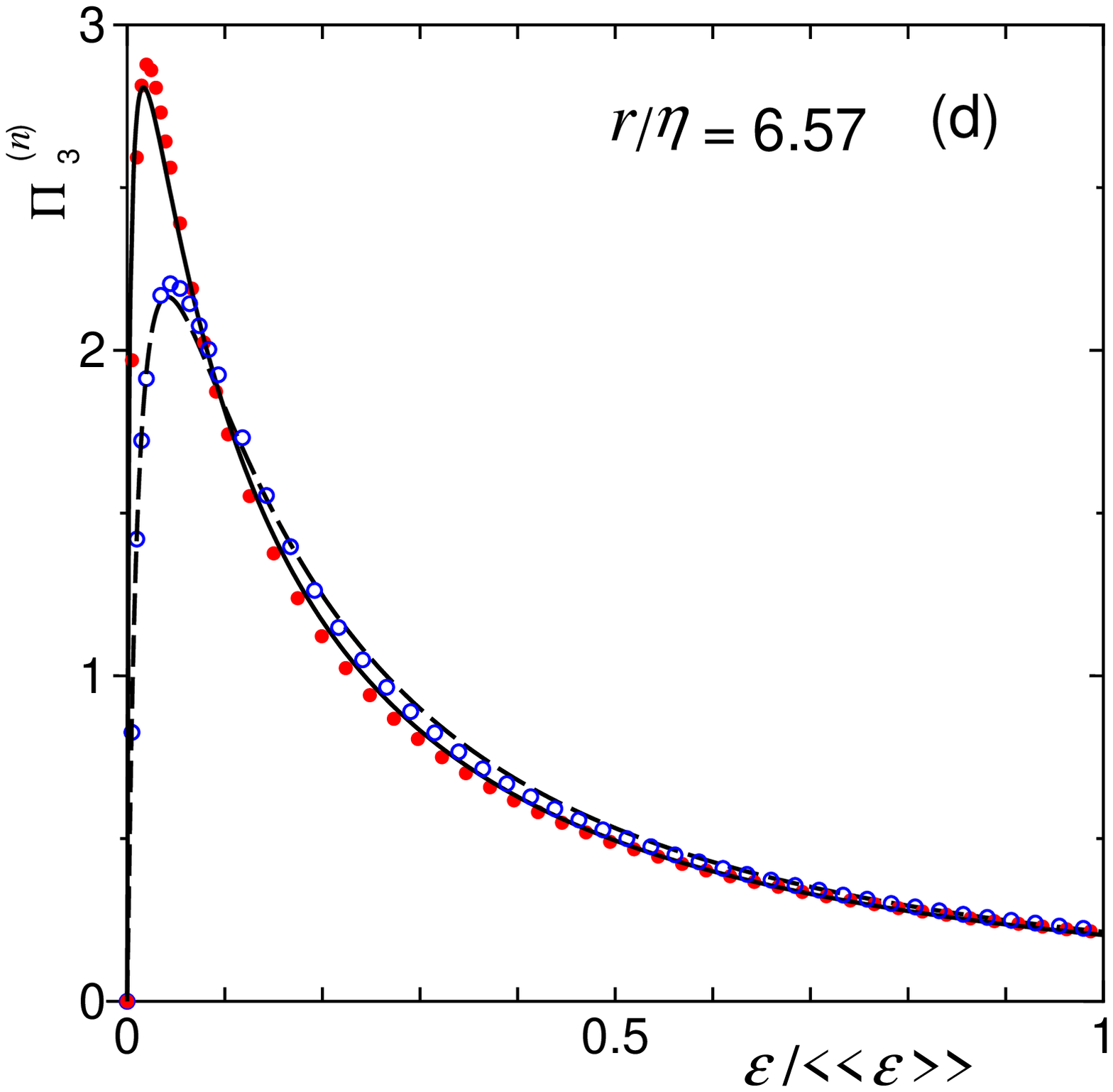}
  \end{center}
 \end{minipage}
 \begin{minipage}{0.33\hsize}
  \begin{center}
   \includegraphics[width=49mm]{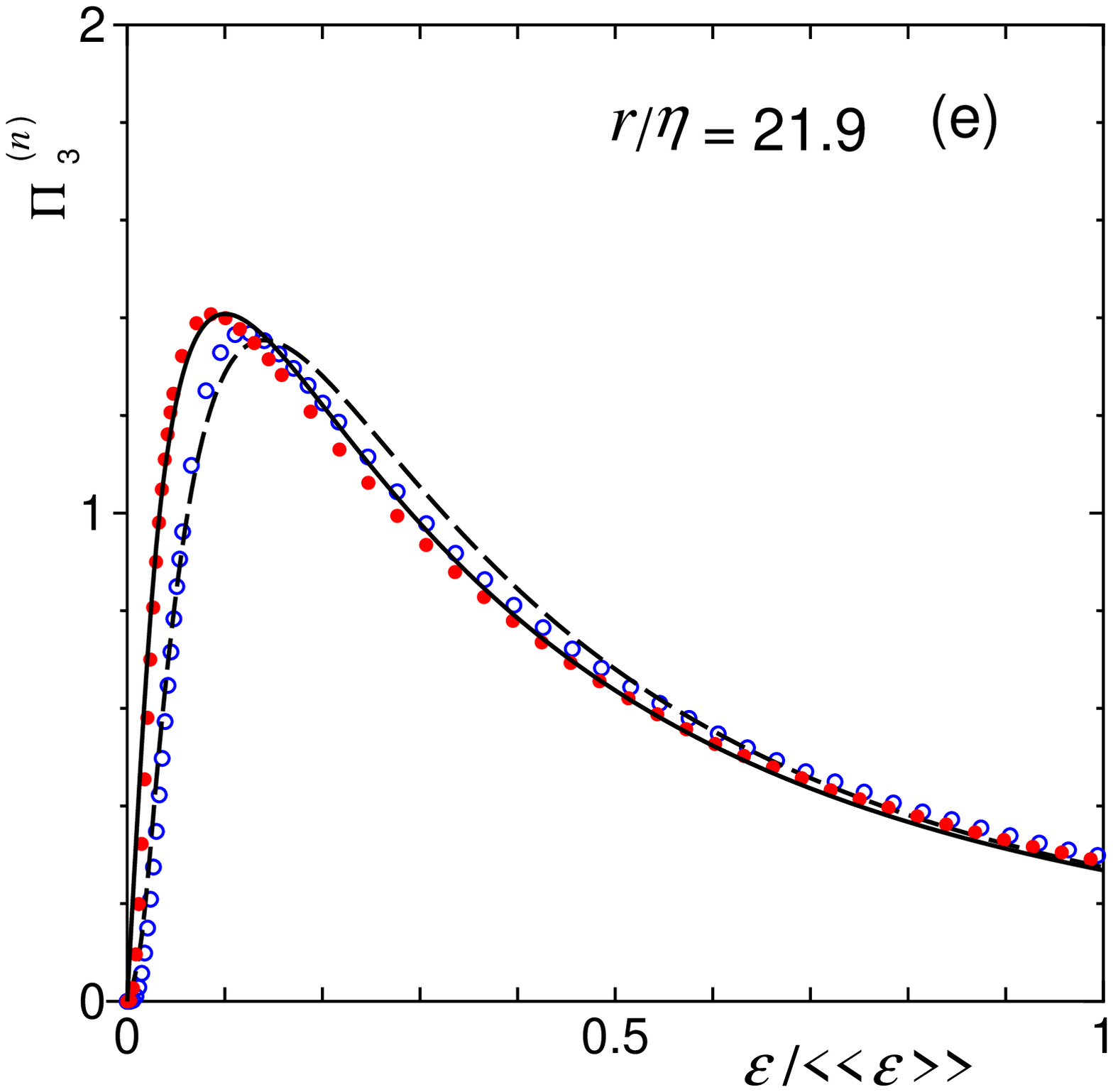}
  \end{center}
 \end{minipage}
 \begin{minipage}{0.33\hsize}
  \begin{center}
   \includegraphics[width=49mm]{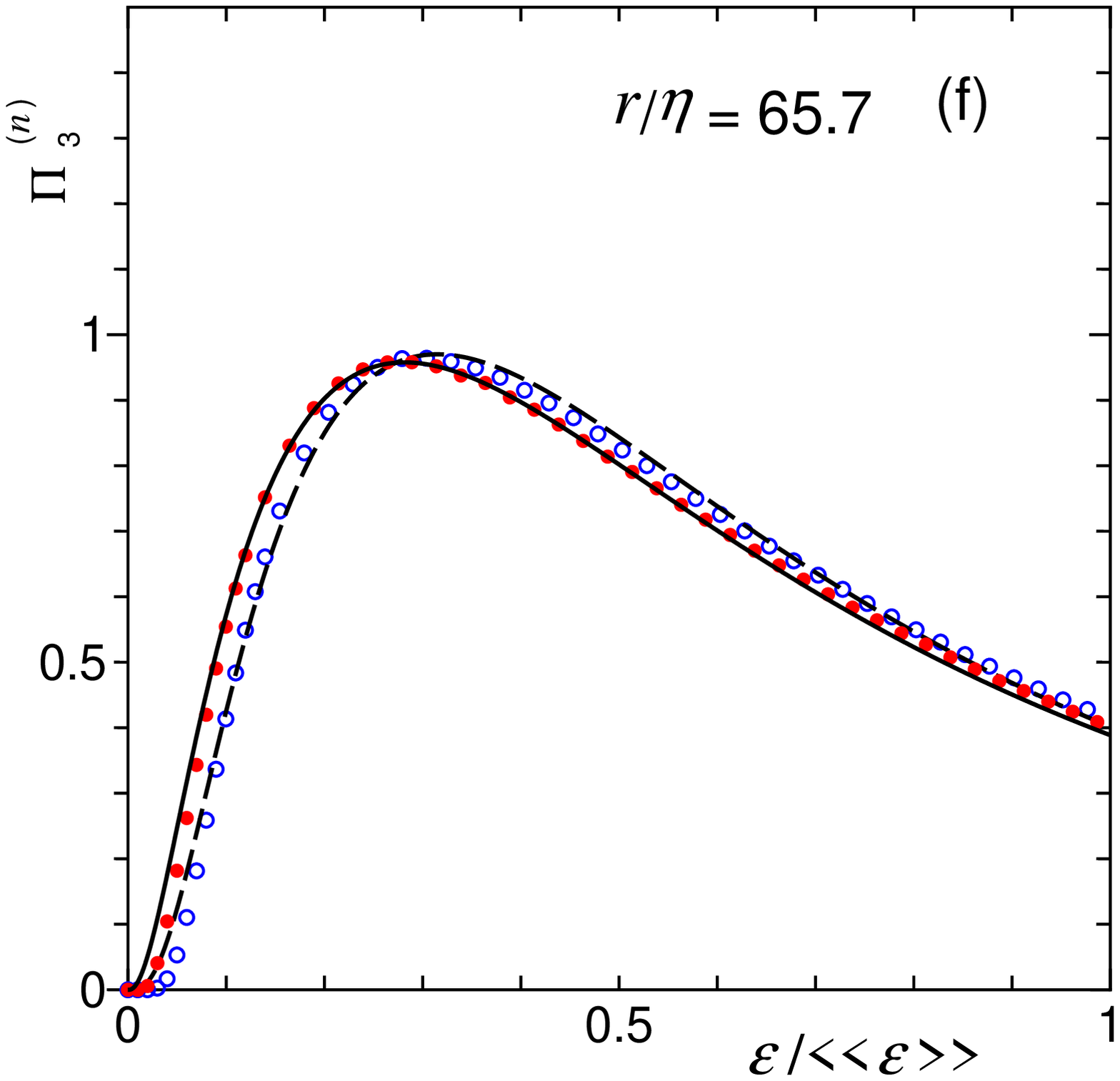}
  \end{center}
 \end{minipage}
\caption{Comparison of PDFs for energy dissipation rates
$\Pi_3^{(n)} (\varepsilon/\dbra \varepsilon \dket)$ and 
$\Pi_3^{(n)} (\varepsilon^{(0)}/\dbra \varepsilon \dket)$ 
created, respectively, with the formulae (\ref{vel der correct up to 4th order}) 
and (\ref{vel der correct up to 1st order}). 
In (a) and (d)--(f), closed (open) circles and full (dashed) lines
are, respectively, the experimental and theoretical PDFs 
for $\Pi_3^{(n)} (\varepsilon/\dbra \varepsilon \dket)$
($\Pi_3^{(n)} (\varepsilon^{(0)}/\dbra \varepsilon \dket)$) with 
$\mu = 0.260$. 
PDFs in (a) represent for the cases $r/\eta = 6.57$ (top), $21.9$ (middle) 
and $65.7$ (bottom), shifted by $-2$ unit along the vertical axis
for better visibility.
The magnification of the central part PDFs for each $r$ ($=\ell_n$) is given in (d)--(f).
The relative difference 
$\Delta_n = [\Pi_3^{(n)} (\varepsilon^{(0)}/\dbra \varepsilon \dket)
- \Pi_3^{(n)} (\varepsilon/\dbra \varepsilon \dket)]/\Pi_3^{(n)} (\varepsilon/\dbra \varepsilon \dket)$ 
is given for (b) $r/\eta = 6.57$ and (c) 65.7, in which closed circles (full lines) are 
experimental (theoretical) $\Delta_n$. 
The error bar is the standard deviation of 100 hidden (not appeared in the figures) data points 
for $\Delta_n$ which locate between the adjacent data points for $\Delta_n$ appeared in the figures.
The shown error bars are thinned out.
Note that (a) and (b)--(f) are, respectively, drawn on log and linear scale in the vertical axes. 
\label{fig:pdf edr delta_3 comp 1st}}
\end{figure*}

We compare, in figures~\ref{fig:pdf edr delta_3 comp 1st} (a) and (d)--(f), 
the PDFs of energy dissipation rates $\Pi_3^{(n)} (\varepsilon/\dbra \varepsilon \dket)$ 
and $\Pi_3^{(n)} (\varepsilon^{(0)}/\dbra \varepsilon \dket)$ which are created, respectively, 
with the help of formulae (\ref{vel der correct up to 4th order}) 
and (\ref{vel der correct up to 1st order}). 
Note that the arguments for every PDFs are scaled by $\dbra \varepsilon \dket$
which does not depend on $r$ ($= \ell_n$).
In figure~\ref{fig:pdf edr delta_3 comp 1st}~(a) each PDF is displayed 
on log scale in vertical axis for the cases $r/\eta = 6.57$ (top), 
$21.9$ (middle) and $65.7$ (bottom), which
are shifted by $-2$ unit along the vertical axis for better visibility.
The magnification of their central part PDFs are displayed
in figures~\ref{fig:pdf edr delta_3 comp 1st}~(d) $r/\eta = 6.57$, (e) $21.9$ and (f) $65.7$
on linear scale in vertical axis.
The closed (open) circles and the full (dashed) lines
are, respectively, the experimental and theoretical PDFs 
for $\Pi_3^{(n)} (\varepsilon/\dbra \varepsilon \dket)$
($\Pi_3^{(n)} (\varepsilon^{(0)}/\dbra \varepsilon \dket)$) with 
$\mu = 0.260$. 
Note that the values of the intermittency exponent $\mu$ for 
$\Pi_3^{(n)} (\varepsilon/\dbra \varepsilon \dket)$ and for
$\Pi_3^{(n)} (\varepsilon^{(0)}/\dbra \varepsilon \dket)$ turn out to be the same.  
Other parameters are listed in table~\ref{pdf edr parameters4/1} and 
table~\ref{pdf edr connection points4/1}.

The relative differences 
$\Delta_n = [\Pi_3^{(n)} (\varepsilon^{(0)}/\dbra \varepsilon \dket)
- \Pi_3^{(n)} (\varepsilon/\dbra \varepsilon \dket)]/\Pi_3^{(n)} (\varepsilon/\dbra \varepsilon \dket)$ 
for $r/\eta =$ $6.57$ and $65.7$ are given, respectively, 
in figures~\ref{fig:pdf edr delta_3 comp 1st}~(b) and (c), in which 
closed circles (full lines) represent experimental (theoretical) mean values of $\Delta_n$. 
The error bar in these figures is the standard deviation of 100 hidden 
(not appeared in the figures) data points for $\Delta_n$ which locate between 
the adjacent data points for $\Delta_n$ appeared in the figures.
These figures show that the mean relative difference $\Delta_n$ in the region of central part of PDFs 
is about 10 times larger than the mean relative difference in the region of tail part.\footnote{
Note that the connection points of $\Pi_3^{(n)} (\varepsilon/\dbra \varepsilon \dket)$
($\Pi_3^{(n)} (\varepsilon^{(0)}/\dbra \varepsilon \dket)$) for 
$r/\eta = 6.57$ and $65.7$ locate, respectively, at 
$\varepsilon^*/\dbra \varepsilon \dket = 4.06$ ($\varepsilon^{(0) *}/\dbra \varepsilon \dket = 4.17$) 
and $\varepsilon^*/\dbra \varepsilon \dket = 5.20$ 
($\varepsilon^{(0) *}/\dbra \varepsilon \dket = 5.20$). 
}
The small negative but nearly constant mean values of $\Delta_n$ in the tail part region 
tell us that the tail of $\Pi_3^{(n)} (\varepsilon/\dbra \varepsilon \dket)$
and that of $\Pi_3^{(n)} (\varepsilon^{(0)}/\dbra \varepsilon \dket)$ are parallel 
with each other, which gives the reason why we obtained the same $\mu$ value for both PDFs.
We observe that the error bars in the tail part region become larger toward the rightmost end of PDF,
which may be attributed to the smaller number of realizations in each bin there.
Actually, the length of an error bar associated with a bin is quite close to the value 
$\sqrt{1/N + 1/N^{(0)}}$ which estimates the standard deviation of 
$\Pi_3^{(n)} (\varepsilon^{(0)}/\dbra \varepsilon \dket)
/\Pi_3^{(n)} (\varepsilon/\dbra \varepsilon \dket)$
with the help of the number of the realizations $N$ ($N^{(0)}$) in the bin under consideration 
for $\Pi_3^{(n)} (\varepsilon/\dbra \varepsilon \dket)$ 
($\Pi_3^{(n)} (\varepsilon^{(0)}/\dbra \varepsilon \dket)$). 
The numbers of realizations in figure~\ref{fig:pdf edr delta_3 comp 1st} (c) are, for example, 
$N = 21$, $N^{(0)} = 23$ for the rightmost error bar,
$N = 206$, $N^{(0)} = 192$ for the fifth error bar from the rightmost error bar,
$N = 31508$, $N^{(0)} = 31327$ for an error bar at $\varepsilon/\dbra \varepsilon \dket \approx 5$ and
$N = 2861811$, $N^{(0)} = 2886837$ for an error bar at around the peak point of 
$\Pi_3^{(n)} (\varepsilon/\dbra \varepsilon \dket)$ where 
$\varepsilon/\dbra \varepsilon \dket \approx 0.25$.
The $\varepsilon$-dependence of the mean values of $\Delta_n$ in the central region indicates 
that the central part of $\Pi_3^{(n)} (\varepsilon^{(0)}/\dbra \varepsilon \dket)$ around
its peak point moves to rightwards relative to the central part of 
$\Pi_3^{(n)} (\varepsilon/\dbra \varepsilon \dket)$.
On the other hand, from the $\varepsilon$-dependence of the mean values of $\Delta_n$
in the tail region, it may be appropriate to interpret that the tail part of 
$\Pi_3^{(n)} (\varepsilon^{(0)}/\dbra \varepsilon \dket)$ moves to leftwards relative to 
the tail part of $\Pi_3^{(n)} (\varepsilon/\dbra \varepsilon \dket)$. 
If the number of realizations in each bin are increased, i.e., statistics are raised, 
we expect that the standard deviations of $\Delta_n$ should reduce their values 
and that the fluctuation of the mean values of $\Delta_n$ in the tail region should disappear.
The difference of the squared time derivatives of (\ref{vel der correct up to 1st order}) 
and (\ref{vel der correct up to 4th order}) gives
$
(\itDelta^{(0)} v/\itDelta t )^2 - (\itDelta^{(3)} v/\itDelta t )^2
= (\partial v /\partial t ) ( \partial^2 v /\partial t^2 ) \itDelta t
+ {\cal O} (\itDelta t )^4
$.
From the direction of the relative horizontal shift of the PDFs, 
we know that the net contributions of the velocity component $v$ for the region 
around the peak point and of the tail region satisfy, respectively, 
\be
\left(\partial v / \partial t \right)\ \left(\partial^2 v / \partial t^2 \right) > 0
\quad \mbox{and} \quad
\left(\partial v / \partial t \right)\ \left(\partial^2 v / \partial t^2 \right) < 0.
\label{outcome from the direction of shift}
\ee
Taking into account the smallness of the gradient of tail part PDFs, we see that
the absolute value of the latter in (\ref{outcome from the direction of shift}) 
is quite large compared with the former value.

The dependence of $\alpha^*$ and $\theta$ on $r/\eta$ ($= \ell_n/\eta$)
are given, respectively, in figure~\ref{fig:pdf edr n_als_theta relation} (b) and (c) 
by open circles for $\delta = 2$, by open squares for $\delta = 3$
and by open triangles for $\delta = 5$,
which are extracted from the series of PDFs cleated with (\ref{vel der correct up to 1st order}).
The dashed line in each figure, (b) and (c), represents 
an empirical formula obtained from all the data points for $\delta = 2$, 3 and 5
by the method of least squares.
These figures prove again, even for the case of full contamination, the correctness of 
the assumption that the fundamental quantities of turbulence are independent of $\delta$.
We also found that $w_3$ is independent of $\delta$ and has 
a common line 
$
\log_{10} w_3 = 0.318 \log_{10} \left(r / \eta \right) + \log_{10} 0.141
$.
The value of $q^\prime$ is found to be about $q^\prime = 1.05$
(see table~\ref{pdf edr parameters4/1}~(b)).

There is only a slightly visible difference of the lines for $\alpha^*$, $w_3$ 
and of the values $q'$ between those obtained from the two kinds of PDFs, 
one with less contamination (\ref{vel der correct up to 4th order}) 
and the other with full contamination (\ref{vel der correct up to 1st order})
(see figure~\ref{fig:pdf edr n_als_theta relation}~(b);
see also table~\ref{pdf edr parameters4/1} and table~\ref{pdf edr connection points4/1}).
The significant difference appears in the $r/\eta$ dependence of $\theta$ 
which are shown in figure~\ref{fig:pdf edr n_als_theta relation}~(c). 
The difference in $\theta$ explains the shift of the peak points between the two PDFs
(see figures~\ref{fig:pdf edr delta_3 comp 1st}~(d)--(f)).

\section{Summary and Discussion \label{summary}}

The new scaling relation (\ref{new scaling relation}) is essential for the parameters 
$\alpha_0$, $X$ and $q$, associated with the tail part PDF, to be determined 
self-consistently as functions of the intermittency exponent $\mu$, and to be independent of 
the magnification rate $\delta$.
On the other hand, we introduced the trial function (\ref{exp g edr}) for the central part PDF
with three adjustable parameters $q'$, $w_3$ and $\theta$, and found that these parameters 
are also independent of $\delta$,
and satisfy scaling behaviors in wider area not restricted to the inertial range.

The independence of $\tilde{n}$ from $\delta$ ensures
the uniqueness of the PDF of $\alpha$ for any value of $\delta$.
The comparison between the empirical formulae for $\tilde{n}$ given in 
figure~\ref{fig:pdf edr n_als_theta relation}~(a) and the theoretical formula (\ref{ell n tilde})
provides us with the estimation $\ell_0 = 20.6$ cm 
which is about the same as the correlation length $17.9$ cm or the separation $20$ cm
of the axes of adjacent rods forming the grid.
Here, we are assuming that the empirical formulae are effective even for $r/\eta \lesssim 2$
(see the discussion in section~\ref{pdf edr} about the effective region of $r/\eta$).
Note that $\ell_0$ gives an estimation of the diameter of the largest eddy 
within the energy cascade model.

As for the parameters appeared in the trial function for the central part PDFs, 
$\exp[-g(\xi_n)]$ in (\ref{exp g edr}), 
the discoveries that $q' \simeq 1.05$ and that $\theta$ and
$\ln w_3$ reveal scaling properties are quite attractive for the research 
looking for the nature of the fluctuations 
surrounding the coherent turbulent motion of fluid.
The fact that the value $q'$ is quite close to 1 indicates that 
the HCT type function in (\ref{exp g edr}), 
i.e., the part giving $\exp[-g(\xi_n)] (\xi_n^*/\xi_n)^{\theta-1}$,
is close to an exponential function. 
There is no theoretical prediction yet, which is based on an ensemble theoretical aspect 
or on a dynamical aspect starting with the N-S equation, to produce 
the formula for the central part PDF that represents the contributions both 
of the coherent turbulent motion providing 
intermittency and of incoherent fluctuations (background flow) around 
the coherent motion.
If one could succeed to formulate a dynamical theory which produces properly
the formula for the central part of PDFs starting with the N-S equation, 
it may provide us with the physical meaning of the parameters $q'$, $\theta$ and $w_3$, 
and with an appropriate pathway to the dynamical approach,
e.g., the renormalization group approach,
to fully developed turbulence. 
A study to this direction is in progress.

Introducing two difference formulae (\ref{vel der correct up to 4th order}) 
and (\ref{vel der correct up to 1st order}) for the estimate of $\partial v/\partial t$, 
i.e., $\itDelta^{(3)} v /\itDelta t$ with less contamination 
and $\itDelta^{(0)} v /\itDelta t$ with full contamination,
we performed a trial for the extraction of information from PDFs by comparing 
two kinds of PDFs for energy dissipation rates, 
$\Pi_3^{(n)} (\varepsilon/\dbra \varepsilon \dket)$ 
and $\Pi_3^{(n)} (\varepsilon^{(0)}/\dbra \varepsilon \dket)$
with $\varepsilon \propto (\itDelta^{(3)} v /\itDelta t)^2$ and
$\varepsilon^{(0)} \propto (\itDelta^{(0)} v /\itDelta t)^2$.
We observed that the intermittency exponents for the two kinds of PDFs
turn out to take the same value $\mu = 0.260$
(see table~\ref{pdf edr parameters4/1} and table~\ref{pdf edr connection points4/1} 
for other parameters).
Through the accurate analyses of PDFs,
it was also revealed that the parameters for $\Pi_3^{(n)} (\varepsilon/\dbra \varepsilon \dket)$ 
and for $\Pi_3^{(n)} (\varepsilon^{(0)}/\dbra \varepsilon \dket)$
are independent of $\delta$ thanks to the new scaling relation (\ref{new scaling relation}), 
and that they show quite similar scaling behaviors extending to the regions 
with smaller and larger $r/\eta$ values outside the inertial range
(see figure~\ref{fig:pdf edr n_als_theta relation}).
The connection points $\alpha^*$ of the tail and central parts of the PDFs take
almost the same value for each $r/\eta$ 
(see table~\ref{pdf edr connection points4/1} and 
figure~\ref{fig:pdf edr n_als_theta relation}~(b)).
It is found that, among the parameters controlling the central part, only $\theta$ has
a relatively larger deviation between the two different PDFs 
(see table~\ref{pdf edr parameters4/1} and figure~\ref{fig:pdf edr n_als_theta relation}~(c)), 
which is related to the shift of the peak point occurred between the two kinds of PDFs.
Other parameters $q'$ and $w_3$ do not have significant difference among the two PDFs 
(see table~\ref{pdf edr parameters4/1}).

Observing the relative difference $\Delta_n$ between 
$\Pi_3^{(n)} (\varepsilon^{(0)}/\dbra \varepsilon \dket)$ 
and $\Pi_3^{(n)} (\varepsilon/\dbra \varepsilon \dket)$ in 
figures~\ref{fig:pdf edr delta_3 comp 1st} (b) and (c) with 
the values $\varepsilon_n^* / \dbra \varepsilon_n \dket = 4.53$ for 
$r/\eta = 6.57$ and $\varepsilon_n^* / \dbra \varepsilon_n \dket = 5.20$ for 
$r/\eta = 65.7$, we notice that the connection point $\varepsilon_n^*$ of 
the center part PDF and the tail part PDF provides us with the boundary dividing two regions 
according to their nature of stability specified by the inequalities in
(\ref{outcome from the direction of shift}).
It seems to tell us that the net behavior of incoherent motion of fluid 
contributing mainly around the peak point (central part) of PDF is  
an {\em unstable} time-evolution, whereas that of coherent turbulent motion 
contributing mainly to the tail part of PDF is a {\em stable} time-evolution.
The former may be attributed to a manifestation of fluctuations, 
whereas the latter to the characteristics of intermittency.
Note that we assumed that the central part $\hat{\Pi}_{3,{\rm cr}}^{(n)}(\xi_n)$
is constituted of two contributions, one from the coherent contribution
$\Pi_{3,{\rm S}}^{(n)}(\varepsilon_n)$ 
and the other from the incoherent contribution $\itDelta \Pi_3^{(n)}(\varepsilon_n)$,
and that almost all the contribution to the tail part $\hat{\Pi}_{3,{\rm tl}}^{(n)}(\xi_n)$
comes from the coherent intermittent motion of turbulence.
Further investigation about these outcomes and their interpretation is necessary,
which we leave as one of the attractive future problems.\footnote{
We observed that there is no visible difference 
between $\Pi_3^{(n)} (\varepsilon/\dbra \varepsilon \dket)$ 
and the PDF extracted with the formula
$
\partial v/\partial t \simeq [v(t+ \itDelta t) - v(t -\itDelta t)]/2\itDelta t
$
which is correct without contamination up to the term of ${\cal O}(\itDelta t)$.
}

Let us close this paper by noting the studies in progress which are deeply related to
the present work.
It has been revealed (Motoike and Arimitsu 2012) that the new scaling relation 
(\ref{new scaling relation}) is intimately related to the $\delta^\infty$ periodic orbits 
($\delta \geq 3$) located at the threshold to chaos via 
the $\delta$ ($\geq 3$) times ramification (bifurcation) 
in $\delta$-period windows in the chaotic region, for example, of the logistic map.
The self-similar nesting structure of $\delta^k$-period windows ($k=1,2,3,\cdots$)
can be an origin of intermittent coherent motion in fully developed turbulence.
We expect that further investigation to this direction to extract a message provided 
by the new scaling relation may lead us to a novel interpretation of 
fully developed turbulence.
We are also performing a precise comparison between the results extracted in this paper 
for the grid turbulence in a wind tunnel and those for 4096$^3$ DNS turbulence 
by raising the resolution of PDFs, i.e., by creating more data points for PDFs.
The results of these studies will be published elsewhere in the near future.

\ack
The authors (T.A. and N.A.) would like to thank Prof.~T.~Motoike,
Dr.~K.~Yoshida, Mr.~M.~Komatsuzaki and Mr.~K.~Takechi for fruitful discussion.

\section*{References}

\begin{harvard}
\item[] Aoyama T, Ishihara T, Kaneda Y, Yokokawa M, Itakura K and Uno A 2005 
	Statistics of energy transfer in high-resolution direct numerical simulation of 
	turbulence in a periodic box 
	{\it J.\ Phys.\ Soc.\ Jpn.} {\bf 74} 3202--3212
\item[] Arimitsu N and Arimitsu T 2002 
	Multifractal analysis of turbulence by using statistics based on non-extensive 
	Tsallis' or extensive R\'enyi's entropy
	{\it J. Korean Phys. Soc.} {\bf 40} 1032--1036
\item[] Arimitsu N and Arimitsu T 2011 
	Verification of the scaling relation within MPDFT by analyzing PDFs for energy 
	dissipation rates out of 4096$^3$ DNS 
	{\it Physica \rm A} {\bf 390} 161--176 
\item[] Arimitsu T and Arimitsu N 2000a 
	Analysis of Fully Developed Turbulence in terms of Tsallis Statistics 
	{\it Phys. Rev.\ \rm E} {\bf 61} 3237--3240
\item[] Arimitsu T and Arimitsu N 2000b 
	Tsallis statistics and fully developed turbulence 
	{\it J.\ Phys.\ {\rm A}: Math.\ Gen.} {\bf 33} L235--L241 
	[{\footnotesize CORRIGENDUM}: 2001 {\bf 34} 673--674]
\item[] Arimitsu T and Arimitsu N 2001 
	Analysis of turbulence by statistics based on generalized entropies 
	{\it Physica \rm A} {\bf 295} 177--194
\item[] Arimitsu T and Arimitsu N 2002 
	PDF of velocity fluctuation in turbulence by a statistics based on generalized entropy 
	{\it Physica \rm A} {\bf 305} 218--226
\item[] Arimitsu T and Arimitsu N 2011 
	Analysis of PDFs for energy transfer rates from 4096$^3$ DNS 
	---Verification of the scaling relation within MPDFT 
	{\it J.\ Turbulence} {\bf 12} 1--25 
\item[] Benzi R, Paladin G, Parisi G and Vulpiani A 1984 
	On the multifractal nature of fully developed turbulence and chaotic systems 
	{\it J.\ Phys.\ {\rm A}: Math.\ Gen.} {\bf 17} 3521-3531
\item[] Benzi R, Biferale L, Paladin G, Vulpiani A and Vergassola M 1991 
	Multifractality in the statistics of the velocity gradients in turbulence 
	{\it Phys.\ Rev.\ Lett.} {\bf 67} 2299--2302
\item[] Biferale L, Boffetta G, Celani A, Devenish B J, Lanotte A and Toschi F 2004 
	Multifractal statistics of Lagrangian velocity and acceleration in turbulence 
	{\it Phys.\ Rev.\ Lett.} {\bf 93}  064502-1-4
\item[] Biferale L and Procaccia I 2005 Anisotropy in Turbulent Flows and in Turbulent Transport
	{\it Phys.\ Rep.} {\bf 414} 43--164 
\item[] Chevillard L, Castaing  B, L\'ev\^eque E and Arneodo A 2006 
	Unified multifractal description of velocity increments statistics in turbulence: 
	Intermittency and skewness 
	{\it Physica \rm D} {\bf 218} 77--82
\item[] Cleve J, Greiner M and Sreenivasan K R 2003
	On the effects of surrogacy of energy dissipation in determining the intermittency exponent
	in fully developed turbulence
	{\it Europhys.\ Lett.} {\bf 61} 756--761
\item[] Costa U M S, Lyra M L, Plastino A R and Tsallis C 1997 
	Power-law sensitivity to initial conditions within a logistic-like family of maps: 
	Fractality and nonextensivity 
	{\it Phys.\ Rev.\ \rm E} {\bf 56} 245--250
\item[] Dubrulle B 1994 
	Intermittency in fully developed turbulence: log-Poisson statistics and 
	generalized scale covariance
	{\it Phys.\ Rev.\ Lett.}  {\bf 73} 959--962
\item[] Grassberger P 1983 
	Generalized dimension of strange attractors 
	{\it Phys.\ Rev.\ Lett.\ \rm A} {\bf 97} 227--229
\item[] Halsey T C, Jensen M H, Kadanoff L P, Procaccia I and Shraiman B I 1986 
	Fractal measures and their singularities: The characterization of strange sets 
	{\it Phys.\ Rev.\ \rm A} {\bf 33} 1141--1151 
\item[] Havrda J H and Charvat F 1967 
	Quantification methods of classification processes: Concepts of 
	structural $\alpha$ entropy 
	{\it Kybernatica} {\bf 3} 30--35 
\item[] Hentschel H G E and Procaccia I 1983 
	The infinite number of generalized dimensions of fractals and strange attractors 
	{\it Physica \rm D} {\bf 8} 435--444 
\item[] Hosokawa I 1991 
	Turbulence models and probability distributions of dissipation and relevant 
	quantities in isotropic turbulence 
	{\it Phys.\ Rev.\ Lett.} {\bf 66} 1054--1057
\item[] Lyra M L and Tsallis C 1998 
	Nonextensivity and multifractality in low-dimensional dissipative systems 
	{\it Phys.\ Rev.\ Lett.} {\bf 80} 53--56 
\item[] Mandelbrot B B 1974 
	Intermittent turbulence in self-similar cascades: Divergence of high moments 
	and dimension of the carrier 
	{\it J.\ Fluid Mech.} {\bf 62} 331--358 
\item[] Meneveau C and Sreenivasan K R 1987 
	The multifractal spectrum of the dissipation field in turbulent flows 
	{\it Nucl.\ Phys.\ \rm B ({\it Proc.\ Suppl.})} {\bf 2} 49--76 
\item[] Motoike T and Arimitsu T 2012
	in preparation to submit

\item[] Mouri H, Hori A and Takaoka M 2008 
	Fluctuations of statistics among subregions of a turbulence velocity field 
	{\it Phys.\ Fluids} {\bf 20} 035108-1--6
\item[] Nelkin M 1990 
	Multifractal scaling of velocity derivatives in turbulence 
	{\it Phys.\ Rev.\ \rm A} {\bf 42} 7226-7229
\item[] Parisi G and Frisch U 1985 
	{\it Turbulence and predictability in geophysical fluid dynamics and climate dynamics} 
	(New York: North-Holland/American Elsevier) pp~84-87
\item[] R\'{e}nyi A 1961 
	On measures of entropy and information 
	{\it Proc. of the 4th Berkeley Symp. on Mathematical Statistics and Probability} 
	(Berkeley: USA) pp~547--561 
\item[] She Z-S and Leveque E 1994 
	Universal scaling laws in fully developed turbulence 
	{\it Phys.\ Rev.\ Lett.} {\bf 72} 336--339
\item[] She Z-S and Waymire E C 1995 
	Quantized energy cascade and log-Poisson statistics in fully developed turbulence
	{\it Phys.\ Rev.\ Lett.} {\bf 74} 262--265
\item[] Suyari H and Wada T 2006 
	Scaling property and Tsallis entropy derived from a fundamental nonlinear 
	differential equation 
	{\it Proc.\ of the 2006 Int.\ Symp.\ on Inf.\ Theory and its Appl.} 
	(ISITA2006) pp~75--80 ({\it Preprint} cond-mat/0608007)
\item[] Tsallis C 1988 
	Possible generation of Boltzmann-Gibbs statistics 
	{\it J.\ Stat.\ Phys.} {\bf 52} 479--487
\item[] Tsallis C 2001 
	Nonextensive statistical mechanics and thermodynamics: Historical background and 
	present status {\it Nonextensive Statistical Mechanics and Its Applications} 
	ed S Abe and Y Okamoto (Berlin: Springer-Verlag) pp~3--98

\end{harvard}

\end{document}